\documentclass[pra, longbibliography,floatfix,superscriptaddress,reprint,]{revtex4-2}
\usepackage{datetime2}

\usepackage{amsmath}
\usepackage{amsfonts}
\usepackage{mathtools}
\usepackage{mathrsfs}
\usepackage{nicefrac}

\usepackage{graphicx}
\usepackage[colorlinks=true,
    citecolor=green,
    linkcolor=red,
    filecolor=magenta,
    urlcolor=blue,]
{hyperref}

\usepackage{physics}

\usepackage{booktabs}

\usepackage{dsfont}
\usepackage{bbm}

\usepackage[capitalize]{cleveref}
\usepackage[dvipsnames]{xcolor}

\usepackage[normalem]{ulem}
\usepackage{soul}

\def \addCQuIC {Center for Quantum Information and Control, University
  of New Mexico, Albuquerque, NM, USA}

\def \addCQuIC {Center for Quantum Information and Control, University of New Mexico, Albuquerque, NM, USA}

\def \addPandAUNM {Department of Physics and Astronomy, University of New Mexico, Albuquerque, NM, USA}
\def \addPandAUO {Department of Physics and Astronomy, University of Oklahoma, Norman, OK, USA}
\def \addLANL {Los Alamos National Laboratory, Los Alamos, NM, USA}

\begin{document}

\title{Neutral atom entanglement using adiabatic Rydberg dressing}

\author{Anupam Mitra}
\email{anupam@unm.edu}
\affiliation{\addCQuIC} \affiliation{\addPandAUNM}

\author{Sivaprasad  Omanakuttan}
\email{somanakuttan@unm.edu}
\affiliation{\addCQuIC} \affiliation{\addPandAUNM}

\author{Michael J. Martin}
\email{mmartin@lanl.gov}
\affiliation{\addLANL}\affiliation{\addCQuIC}

\author{Grant W. Biedermann}
\email{biedermann@ou.edu}
\affiliation{\addPandAUO}

\author{Ivan H. Deutsch}
\email{ideutsch@unm.edu}
\affiliation{\addCQuIC} \affiliation{\addPandAUNM}

\date{\DTMNow}

\begin{abstract}
We revisit the implementation of a two-qubit entangling gate, the M{\o}lmer-S{\o}rensen gate, using the adiabatic Rydberg dressing paradigm for neutral atoms as studied in \cite{mitra2020robust}. We study the implementation of rapid adiabatic passage using a two-photon transition, which does not require the use of an ultra-violet laser, and can be implemented using only amplitude modulation of one field with all laser frequencies fixed. We find that entangling gate fidelities, comparable to the one-photon excitation, are achievable with the two-photon excitation. Moreover, we address how the adiabatic dressing protocol can be used to implement entangling gates outside the regime of a perfect Rydberg blockade. We show that using adiabatic dressing we can achieve scaling of gate fidelity set by the fundamental limits to entanglement generated by the Rydberg interactions while simultaneously retaining a limited population in the doubly-excited Rydberg state. This allows for fast high fidelity gates for atoms separated beyond the blockade radius.

\end{abstract}

\maketitle




\section{Introduction}
Optically trapped arrays of neutral atoms with tunable electric dipole-dipole interactions (EDDI) are a promising platform for scalable quantum computation  \cite{brennen1999quantum, isenhower2010demonstration, saffman2010quantum, saffman2016quantum, levine2019parallel, bluvstein2022quantum, graham2022multi, ebadi2022quantum}, quantum simulations  \cite{zeiher2016many, zeiher2017coherent, borish2020transverse, browaeys2020many, bernien2017probing, keesling2019quantum, ebadi2021quantum, scholl2021quantum}, and quantum metrology  \cite{kaubruegger2019variational, van2021impacts, schine2022long}. 
A variety of protocols have been studied to create entanglement between atomic qubits using the strong EDDI of Rydberg atoms \cite{jaksch2000fast, beterov2013quantum, keating2015robust, beterov2016two, beterov2018adiabatic, levine2019parallel, mitra2020robust, saffman2020symmetric, beterov2020application}, and have been demonstrated in alkali atoms including cesium and rubidium  \cite{zhang2010deterministic, isenhower2010demonstration, wilk2010entanglement, levine2018high, omran2019generation, levine2019parallel, graham2019rydberg, jo2020rydberg, martin2021molmer, bluvstein2022quantum, graham2022multi} and in alkaline earth atoms including strontium and ytterbium  \cite{madjarov2019strontium, schine2022long, ma2021universal}. 
Given rapid advances in the field, we seek to revisit some practical considerations and fundamental limits for qubit entanglement that are achievable with adiabatic Rydberg dressing of ground state atoms.

In particular, we consider the use of adiabatic Rydberg dressing \cite{johnson2010interactions, keating2015robust, mitra2020robust, martin2021molmer, schine2022long}, a powerful tool for robustly creating entanglement in atomic-clock qubits. In this approach, the Rydberg character is adiabatically admixed into one of the clock states through a chirp of the laser frequency and/or intensity ramp \cite{mitra2020robust, martin2021molmer, schine2022long}. The resulting light shift of the dressed state is then mediated by the Rydberg EDDI, leading to entanglement \cite{mitra2020robust, martin2021molmer, schine2022long}.
This tool has been implemented to create Bell states of clock qubits in the microwave \cite{jau2016entangling} and optical regimes \cite{schine2022long} and for studies of many-body physics \cite{zeiher2016many, zeiher2017coherent, borish2020transverse}. Schemes for implementing two-qubit entangling quantum logic gates based on adiabatic Rydberg dressing have been studied theoretically \cite{keating2015robust, mitra2020robust} and recently demonstrated \cite{martin2021molmer, schine2022long}.

Adiabatic Rydberg dressing is most naturally implemented using a one-photon transition between a clock state and a high-lying Rydberg state \cite{keating2015robust, jau2016entangling, zeiher2016many, zeiher2017coherent, borish2020transverse, mitra2020robust}.
Such an approach requires a high-power ultraviolet-laser which is technically challenging and can lead to adverse effects, such as photoelectric charging of dielectrics and spurious electric fields. Adiabatic Rydberg dressing would be more simply achieved through a standard two-photon transition that is typically used for Rydberg excitation, but this may lead to other challenges due to additional decoherence and spurious light shifts from off-resonant excitation to the intermediate state \cite{zhang2012fidelity, de2018analysis}. We revisit this problem here and show that a two-photon excitation is well-matched to adiabatic Rydberg dressing, with additional light shifts facilitating adiabatic passages by modulating only one laser amplitude. With the current state of the art, decoherence will not greatly reduce gate fidelity. Moreover, dominant inhomogeneities can be removed in this protocol through spin echoes, as studied in \cite{mitra2020robust} implemented in  \cite{martin2021molmer, schine2022long}.

Beyond the practical consideration of two-photon excitation for adiabatic Rydberg dressing, we revisit the fundamental limits of gate fidelity that can be generated using adiabatic Rydberg dressing of ground state atoms. While the basic entangling interaction is due to the EDDI with strength $|V|$, in protocols that employ the Rydberg blockade, the speed of the gate is limited by the effective Rabi frequency of the coupling laser $\Omega_{\mathrm{eff}}$, as in the seminal work of \cite{jaksch2000fast}. Rydberg dressing under a strong blockade, where the admixture of the doubly excited Rydberg states is small and often negligible requires $\hbar \Omega_{\mathrm{eff}} \ll |V|$. As such, one cannot achieve the fundamental scaling in the gate error rate set by the ratio $2\pi\hbar \Gamma/|V|$ for a characteristic decoherence rate $\Gamma$ \cite{saffman2016quantum}. Adiabatic Rydberg dressing has generally also operated in the strong blockade regime \cite{keating2013adiabatic, keating2015robust, mitra2020robust, martin2021molmer, schine2022long}, but this is not essential to the protocol. In principle, adiabatic admixtures that include doubly excited Rydberg levels will strongly increase the entangling energy or may be used to maintain atoms separated beyond the blockade radius where they can be more easily individually addressed, yet still achieve fast gates. Rydberg-mediated entanglement beyond the strong blockade regime has been demonstrated using finely tuned two-atom Rabi oscillations \cite{jo2020rydberg}. In addition, some quantum simulation schemes implementing interacting spin models do not assume strong blockade in a multi-atom array, allowing implementation of elaborate interaction graphs between atoms in one-dimensional  \cite{bernien2017probing, keesling2019quantum, omran2019generation} and two-dimensional geometries  \cite{de2018analysis, ebadi2021quantum, scholl2021quantum, ebadi2022quantum, bluvstein2022quantum, graham2022multi}. 

We show here that by going beyond the perfect blockade regime one can use adiabatic Rydberg dressing to reach the fundamental scaling of entanglement fidelity \cite{wesenberg2007scalable}.
Such an approach may become more feasible, e.g., using bound states of doubly excited Rydberg macrodimers \cite{sassmannshausen2016observation} that have been well resolved \cite{sassmannshausen2016observation, hollerith2021realizing}, and can be employed for such coherent control of entanglement \cite{hollerith2021realizing}. In addition, we find that one can implement entangling gates in the weak blockade regime using an adiabatic Rydberg dressing scheme that requires only a limited population in the doubly-excited Rydberg state, similar to \cite{jo2020rydberg} and unlike some other protocols for entangling gates \cite{levine2018high, levine2019parallel, saffman2010quantum, saffman2016quantum}. Thus, protocols that extend beyond the perfect blockade regime may enable even more powerful schemes for neutral atom quantum information processing. 

The remainder of this article is organized as follows.
In \cref{sec:TwoPhotonAdiabaticPassage} we discuss the implementation of two-photon adiabatic Rydberg dressing passages for creating high fidelity entangling gates.
We show that fidelities $\mathcal{F} > 0.99$ are possible with state-of-the-art experiments. 
In \cref{sec:DressingEntanglingEnergy} we study the scaling of the Rydberg dressing entangling energy in the regimes of strong and weak  blockade and show that we can reach the fundamental scaling as predicted in \cite{wesenberg2007scalable} when we allow a small admixture of doubly excited Rydberg states during adiabatic Rydberg dressing.
In \cref{sec:Conclusions} we conclude and give an outlook toward future applications. 


\section{ Entangling Gates with Adiabatic Dressing}
\label{sec:TwoPhotonAdiabaticPassage}

\begin{figure*}
    \includegraphics[width=0.9\textwidth]    {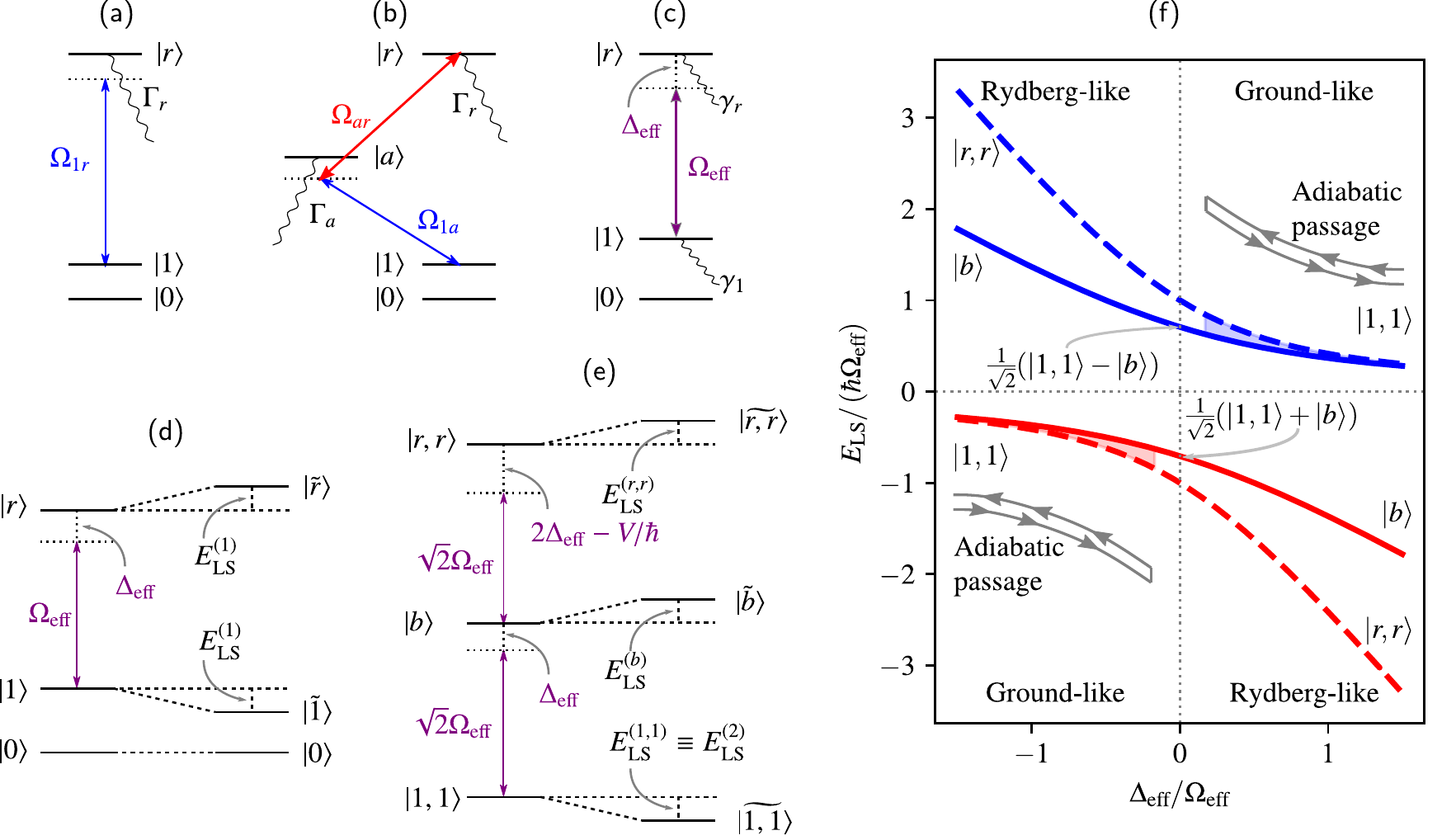}
    \caption{  
    Two-atom energy levels for implementing adiabatic Rydberg dressing. 
    (a) One photon $\ket{1} \leftrightarrow \ket{r}$ transition, with Rabi frequency $\Omega_{1r}$ and Rydberg decay rate $\Gamma_r$.
    (b) Two photon $\ket{1} \leftrightarrow \ket{a} \leftrightarrow \ket{r}$ transition with Rabi frequencies $\Omega_{1a}$ and $\Omega_{ar}$ respectively, intermediate state decay rate $\Gamma_a$ and Rydberg decay rate $\Gamma_r$. 
    (c) Effective three-level system in regime of adiabatically eliminating the intermediate state $\ket{a}$, with effective Rabi frequency $\Omega_{\text{eff}}$ and effective detuning $\Delta_{\text{eff}}$ due to the difference of light-shifts experienced by $\ket{a}$ and $\ket{r}$ and effective decay rate $\gamma_r$ from $\ket{r}$ and $\gamma_1$ from $\ket{1}$.
    (d) Energy levels and lights-shifts in one-atom dressing, where each atom is dressed independently.
    (e) Energy levels and lights-shifts in two-atom dressing, where both atoms are dressed together in the presence of interaction energy $V$.
    (f) Energy shifts of atomic states as a function of detuning, in the strong blockade ($\hbar\Omega_{1r} \ll |V|$, $\hbar|\Delta_{1r}| \ll |V|$) case, which play a role in the adiabatic passage between ground-like states and Rydberg-like states. The shaded region shows the entangling-energy [\cref{eq:EntanglingEnergy}], which is used to accumulate entangling phase.}
    \label{fig:EnergyLevels}
\end{figure*}

We study the implementation of two-qubit gates with qubits encoded in clock states, e.g., $\ket{0} \equiv \ket{(ns),\,  {}^2S_{1/2}, F, m=0}$, $\ket{1} \equiv \ket{(ns),\, {}^2S_{1/2},\, F', m=0}$ for alkali atoms and  $\ket{0} \equiv \ket{(ns)^2,\, {}^1S_0}$,\, $\ket{1} \equiv \ket{(nsnp),\, {}^3P_0}$ for alkaline earth-like atoms. Entanglement is generated by the adiabatic dressing of the $\ket{1}$-state through a one- or two-photon transition to an excited Rydberg state  $\ket{r}$ with high principle quantum number $n_r$. For a one-photon ultraviolet transition, $\ket{r} \equiv \ket{(n_r p),\, {}^2P_J}$ for alkalis and $\ket{r} \equiv \ket{(nsn_r s),\, {}^3S_1}$ for alkaline earths. In the two-photon case, $\ket{r} \equiv \ket{(n_r s),\, {}^2S_{1/2}}$ for alkalis, and $\ket{r} \equiv \ket{(ns n_r p),\, {}^3P_J}$ for alkaline earths, with an intermediate auxiliary state $\ket{a} \equiv \ket{(n_a p),\, {}^2P_J}$ or $\ket{a} \equiv \ket{(ns n_a s),\,{}^3S_1}$ respectively. Generation of entanglement is fundamentally limited by decoherence due to the lifetime of $\ket{r}$ and $\ket{a}$, which depend on the choice of principal quantum numbers $n_r$ and $n_a$. Schematics for one- and two-photon coupling are shown in \cref{fig:EnergyLevels}(a) and \cref{fig:EnergyLevels}(b) respectively.

We consider two atoms symmetrically coupled by uniform laser fields. As only the $\ket{1}$-state is coupled to $\ket{r}$ (in a one- or two-photon transition), the Hamiltonian takes the form
\begin{equation}
    \label{eq:Hamiltonian}
    \hat{H} = \hat{H}_1 \otimes \ketbra{0}{0} +  \ketbra{0}{0} \otimes \hat{H}_1 + \hat{H}_{1,1},
\end{equation}
where $\hat{H}_1$ is the Hamiltonian for one atom in $\ket{1}$ coupled to $\ket{r}$ and $\hat{H}_{1,1}$ is the two-atom coupling, including the Rydberg mediated EDDI. We define the Rabi frequencies $\Omega_{\alpha \beta}$ and detunings $\Delta_{\alpha \beta}$ for each of the corresponding $\ket{\alpha} \leftrightarrow \ket{\beta}$ transitions as shown in \cref{fig:EnergyLevels}(a) and \cref{fig:EnergyLevels}(b). For a two-photon excitation, we consider the regime $\Omega_{1a} \ll |\Delta_{1a}|$ so that the intermediate state can be adiabatically eliminated. In that case, we have the universal single-atom Hamiltonian  
 \begin{equation}
    \hat{H}_1 =
    -\hbar \Delta_\mathrm{eff} \ketbra{r}{r} 
    + \frac{\hbar \Omega_\mathrm{eff}}{2}\left( \ketbra{r}{1}
    +\ketbra{1}{r} \right).
    \label{eq:OneAtomRydbergHamiltonian}
\end{equation}
For the one-photon ultraviolet excitation, $\Omega_\mathrm{eff} = \Omega_{1r}$,  $\Delta_\mathrm{eff} = \Delta_{\mathrm{1r}}$. In the two-photon case $\Omega_\mathrm{eff} = (\Omega_{1a}\Omega_{ar})/(2\Delta_{1a})$ and $ \Delta_\mathrm{eff} = \Delta_{1a}+\Delta_{ar} +(\delta_1 + \delta_r)$, where $\delta_1 = (\Omega^2_{1a})/(4\Delta_{1a})$  and $\delta_r = -(\Omega^2_{ar})/(4\Delta_{ar})$ are the light shifts of levels $\ket{1}$ and $\ket{r}$ respectively due to their coupling to $\ket{a}$. Finally, the entangling two-atom Hamiltonian is
\begin{equation}
\begin{aligned}
    \hat{H}_{1,1} & 
    = \ketbra{1}{1} \otimes \hat{H}_1 +
    \hat{H}_1 \otimes \ketbra{1}{1} + 
    V\ketbra{r,r}{r,r}
    \\ &
    = -\hbar\Delta_\mathrm{eff}
    \qty(\ketbra{b}{b} + \ketbra{d}{d})
    \\ &
    + \left(-2\hbar\Delta_\mathrm{eff} +V\right)
    \ketbra{r,r}{r,r}
    \\ &
    + \frac{\hbar}{2}\sqrt{2} \Omega_\mathrm{eff}
    \left(\ketbra{b}{1,1} + 
    \ketbra{r, r}{b}+ +
    \text{h.c.} \right),
    \label{eq:TwoAtomRydbergHamiltonian}
\end{aligned}
\end{equation}
where $V$ is the atom-atom potential energy arising from the EDDI when both atoms are in $\ket{r}$, 
and $\ket{b} \equiv \qty(\ket{1,r} + \ket{r,1})/\sqrt{2}$, $\ket{d} \equiv \qty(\ket{1,r} - \ket{r,1})/\sqrt{2}$ 
are the bright and dark states, respectively, for symmetric coupling. When $|V|\gg \hbar \Omega_\mathrm{eff},\hbar| \Delta_\mathrm{eff}|$, excitation to the doubly excited Rydberg state is strongly blockaded. In that case we can reduce this Hamiltonian to a two-atom, two-level system 
\begin{equation}
    \label{eq:TwoAtomPerfectBlockade}
    \hat{H}_{1,1} \approx -\hbar\Delta_\mathrm{eff}\ketbra{b}{b}
    + \frac{\hbar}{2}\sqrt{2} \Omega_\mathrm{eff}
    \left(\ketbra{b}{1,1} +  \ketbra{1,1}{b}  \right).
\end{equation}
The effect of the blockade is seen explicitly in the coupling of $\ket{1,1}$ to the entangled bright state $\ket{b}$.

The eigenstates of the Hamiltonian in \cref{eq:Hamiltonian} are the dressed states. In particular, we denote the dressed clock states (computational basis states) $\{ \ket{0,0}, \ket*{0, \tilde{1}}, \ket*{\tilde{1},0}, \ket*{\widetilde{1,1}} \}$. The eigenvalues $E_{0,\tilde{1}} = E_{\tilde{1},0}$ and $E_{\widetilde{1,1}}$ contain contributions from light shifts, $E_{\mathrm{LS}}^{(1)}$ with one atom or $E_{\mathrm{LS}}^{(2)}$ with two atoms coupled to the Rydberg state. The entangling energy, denoted by  $\hbar \kappa$, is the energy difference between the interacting and noninteracting atoms,
\begin{multline}
   \kappa =
   \frac{1}{\hbar} \qty(
   E_{\mathrm{LS}}^{(2)} - 2E_{\mathrm{LS}}^{(1)} )
   \\
   \approx \frac{\Delta_\mathrm{eff}}{2} \pm \frac{1}{2}\left(\sqrt{2 \Omega^2_\mathrm{eff} + \Delta^2_ \mathrm{eff}} - 2\sqrt{ \Omega^2_\mathrm{eff} + \Delta^2 _\mathrm{eff}}\right),
   \label{eq:EntanglingEnergy}
\end{multline}
where the approximation in the second line holds only in the limit of a perfect blockade, with entangling Hamiltonian \cref{eq:TwoAtomPerfectBlockade}, and $\pm$ refers to the two branches of the dressed states in \cref{fig:EnergyLevels}.

\begin{figure}
    \centering
    \includegraphics[width=0.48\textwidth]{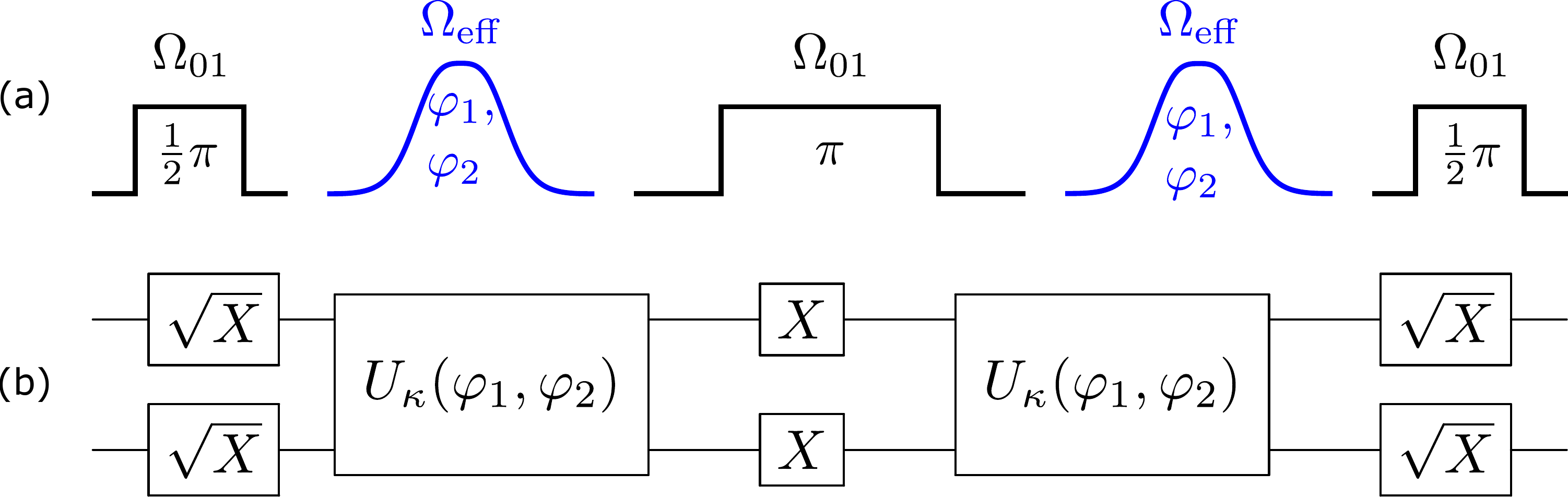}
    \caption{Adiabatic passages interleaved in a spin-echo sequence. (a) Pulse and ramp sequence. (b) Equivalent circuit diagram. When $\varphi_2 =\pi/2$ the result is the $\mathrm{MS}_{yy}$ gate [\cref{eq:MSyyGate}].}
    \label{fig:SpinEchoAdiabaticPassage}
\end{figure}

An entangling gate is achieved through the dynamical phase accumulated from the entangling energy $\varphi_2 = \int\kappa(t') \dd t'$  \cite{keating2015robust, jau2016entangling, lee2017demonstration, mitra2020robust, martin2021molmer, schine2022long}. As discussed in \cite{mitra2020robust}, we consider generating a two-qubit entangling gate using a spin-echo sequence, as shown in \cref{fig:SpinEchoAdiabaticPassage} and demonstrated in \cite{martin2021molmer,schine2022long}. The echo sequence consists of a $\pi/2$ pulse about the $x$-axis, followed by an adiabatic ramp accumulating non-local phase $\varphi_2 = \int \kappa(t') \dd t'$, an echo a $\pi$ pulse about the $x$-axis, followed by another adiabatic ramp accumulating nonlocal phase $\varphi_2 = \int \kappa(t') \dd t'$ , and a final $\pi/2$ pulse about the $x$-axis, as shown in \cref{fig:SpinEchoAdiabaticPassage}(a). An equivalent circuit diagram with the shorthand $\sqrt{X}$ representing a $\pi/2$ pulse about the $x$-axis, $X$ representing a $\pi$ pulse about the $x$-axis and $\hat{U}_{\kappa}(\varphi_1, \varphi_2)$ representing the unitary,
\begin{equation}
\begin{aligned}
   \hat{U}_{\kappa}(\varphi_1, \varphi_2)
   & = 
   \exp(- i \varphi_2 \qty(\frac{\hat{\sigma}_z}{2} \otimes \frac{\hat{\sigma}_z}{2}))
   \\ & \times  
    \exp(- i \varphi_1 \qty(\mathbbm{1} \otimes \frac{\hat{\sigma}_z}{2} + \frac{\hat{\sigma}_z}{2} \otimes \mathbbm{1})),
    \label{eq:UnitaryAdiabaticPassage}
\end{aligned}
\end{equation}
implemented during each adiabatic ramp, is shown in \cref{fig:SpinEchoAdiabaticPassage}(b). Importantly, the spin-echo removes all phases, $\varphi_1$, arising for single atom-light shifts, including the  dominant errors arising from atom thermal motion, and the resulting inhomogeneities  \cite{mitra2020robust, martin2021molmer, schine2022long}. Designing the adiabatic ramps such that $\varphi_2 = \pi/2$ in each ramp, the resulting unitary transformation is a M{\o}lmer-S{\o}rensen YY-gate ($\mathrm{MS}_{yy}$),
 \begin{equation}
     \hat{U}_{\mathrm{MS}_{yy}} = 
     \exp(-\frac{i\pi}{4}\hat{\sigma}_y \otimes \hat{\sigma}_y),
     \label{eq:MSyyGate}
 \end{equation}
 which is a perfect entangler for the qubits, that is a gate that can output maximally entangled states from input product states \cite{nielsen2003quantum, zhang2003geometric, mitra2020robust}. This robust protocol extends to two-photon excitation. Off-resonant coupling to the intermediate state leads to additional light shifts and potential noise due to intensity fluctuations. The spin echo removes this noise in its contribution to the single-atom light shift. Residual noise from this fluctuating light shift results only from the uncertainty in $\Delta_\mathrm{eff}$ and $\Omega_\mathrm{eff}$ and its effect on $\kappa$.
 
The fundamental source of decoherence is due to the decay of the Rydberg state at rate $\Gamma_r$ and the intermediate state at rate $\Gamma_a$. To good approximation the decays will lead to leakage outside the qubit subspace. In that case we can treat decoherence simply through a non-trace-preserving Schr\"{o}dinger evolution with a non-Hermitian Hamiltonian $\hat{H}_\mathrm{eff} = \hat{H} -\frac{i \hbar}{2}\sum_\mu \hat{L}_\mu^\dag \hat{L}_\mu $, where $\{\hat{L}_\mu\}$ are the Lindblad jump operators. In the one-photon excitation, $\sum_\mu \hat{L}_\mu^\dag \hat{L}_\mu = \Gamma_r \ketbra{r}{r}$ for each atom. In the two-photon excitation,
\begin{equation}
    \sum_\mu \hat{L}_\mu^\dag \hat{L}_\mu = 
    \gamma_1 \ketbra{1}{1} +
    \gamma_r \ketbra{r}{r} + 
    \gamma_{1r} \left(\ketbra{r}{1} + \ketbra{1}{r}\right),
\end{equation}
for each atom. Here levels $\ket{1}$ and $\ket{r}$ and their coherences decay due to off-resonant photon scattering with rates 
\begin{equation}
    \gamma_1 = \frac{\Omega_{1a}^2}{4\Delta_{1a}^2}\Gamma_a, \; \gamma_r = \frac{\Omega_{ar}^2}{4\Delta_{ar}^2}\Gamma_a+\Gamma_r, \; \gamma_{1r} = \frac{\Omega_{ra}\Omega_{1a}}{4\Delta_{1a}^2}\Gamma_a.
\end{equation}
High-fidelity gates for two-photon excitation require sufficiently long lifetimes of level $\ket{a}$.

As studied in \cite{mitra2020robust}, the highest fidelity gates are achieved for strong dressing, with the exciting laser close to Rydberg resonance, and a large admixture of $\ket{b}$ in the dressed state $\ket*{\widetilde{1,1}}$. For a one-photon transition, we consider an adiabatic sweep involving a Gaussian laser intensity sweep and the linear detuning sweep, according to,
\begin{widetext}
\begin{equation}
    \begin{aligned}
        &
        \lvert\Delta_{1r}(t) \rvert 
        = 
        \begin{cases}
          \Delta_{\mathrm{max}} + 
          \frac{\Delta_{\mathrm{max}}-\Delta_\mathrm{min}}
          {t_2-t_1} \times (t-t_1),
          & t_1 \leq t < t_2
          \\
          \Delta_{\mathrm{min}},
          & t_2 \leq t \leq t_3 
          \\
          \Delta_{\mathrm{min}} + 
          \frac{\Delta_{\mathrm{min}}-\Delta_\mathrm{max}}
          {t_4-t_3} \times (t-t_3),
          & t_3 < t \leq t_4
        \end{cases}
        \\ &
        \Omega_{1r}(t) =
        \begin{cases}
        \Omega_{\mathrm{min}} + 
         \left(\Omega_\mathrm{max}-\Omega_{\mathrm{min}}\right)\exp\left(-\frac{(t - t_1)^2}{2t_w^2}\right),
          & t_1 \leq t < t_2
          \\
          \Omega_{\mathrm{max}},
          & t_2 \leq t \leq t_3 
          \\
          \Omega_{\mathrm{min}} + 
          \left(\Omega_\mathrm{max} - \Omega_{\mathrm{min}}\right)\exp\left(-\frac{(t - t_3)^2}{2t_w^2}\right),
          & t_3 < t \leq t_4
      \end{cases}.
      \label{eq:GaussianOmega_1rLinearDelta_1r}
    \end{aligned}
\end{equation}
\end{widetext}
The resulting MS-gate was demonstrated in \cite{martin2021molmer, schine2022long}. 

\begin{figure*}
\includegraphics[width=0.99\textwidth]{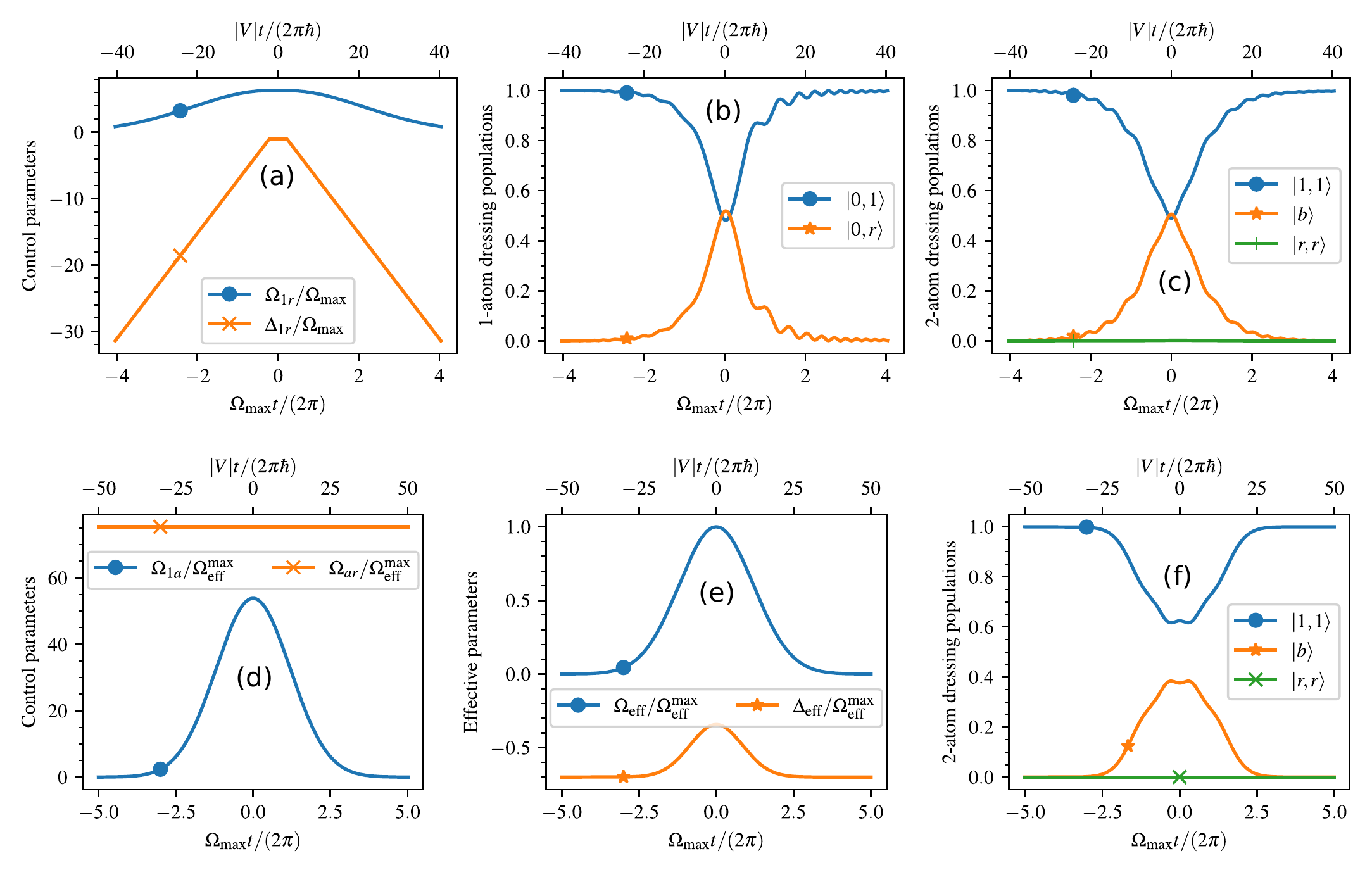}
\caption{ 
Adiabatic passages to implement $\hat{U}_{\kappa}(\varphi_1, \varphi_2)$ with $\varphi_2 = \pi/2$ [\cref{eq:UnitaryAdiabaticPassage}, \cref{fig:SpinEchoAdiabaticPassage}]  in the strong blockade regime ($\hbar \Omega_{\mathrm{eff}} = 0.1 \vert V \vert$). (a) One-photon adiabatic passage Gaussian sweep of Rabi frequency and linear sweep of detuning as in \cite{mitra2020robust, martin2021molmer}. (b) 1-atom populations during a one-photon adiabatic passage  (c) 2-atom population during a one-photon adiabatic passage. (d) Two-photon adiabatic passage using a Gaussian sweep of Rabi frequency $\Omega_{1a}$, with all other parameters fixed, which leads to an effective sweep of the two-photon Rabi frequency $\Omega_{\text{eff}}$ and two-photon detuning $\Delta_{\text{eff}}$ as shown in (e). (f) 2-atom populations during a two-photon adiabatic passage. Bottom axes show time measured in units of $2\pi/\Omega_{\max}$, top axes show time measured in units of $|V|t/(2 \pi \hbar)$. In the strong blockade, as expected, $|V|t/\hbar \gg \Omega_{\max}t$
}
\label{fig:StrongBlockadeAdiabaticPassage}
\end{figure*}
For the two-photon case, the effect of the light shift arising from the intermediate detuning affords additional possibilities for coherent control. We consider the case of exact two-photon resonance in the absence of the light shift, and a fixed Rabi frequency $\Omega_{ar}$ and detuning $\Delta_{ar}$ on the $\ket{a} \leftrightarrow \ket{r}$ transition. Adiabatic dressing is achieved solely through a Gaussian ramp  of the intensity of the laser driving the $\ket{1} \leftrightarrow \ket{a}$ according to the Rabi frequency, 
\begin{equation}
    \Omega_{1a} = 
    \begin{cases}
    \Omega_{1a}^{\mathrm{max}}, \text{ }-\lvert t_{\mathrm{stop}}\rvert \leq t \leq\lvert t_{\mathrm{stop}}\rvert\\
    \Omega_{1a}^{\mathrm{max}} \exp(-\frac{(t-\lvert t_{\mathrm{stop}\rvert})^2}{2 t_w^2}), \text{  otherwise.}\\
    \end{cases}
    \label{eq:GaussianOmega_a}
\end{equation} 
One can modulate $\vert t_{\mathrm{stop}}\vert$,  the time after which the Rabi frequency remains constant, and $t_w$  the width of the Gaussian pulse, to obtain to the desired gate of interest. \cref{fig:StrongBlockadeAdiabaticPassage} shows an example of ramps for the two-photon adiabatic passage as well the population as a function of time during the pulse sequence. 

As discussed above, to implement the M{\o}lmer-S{\o}rensen gate we consider two adiabatic ramps intertwined by the spin echo sequence as shown in \cref{fig:SpinEchoAdiabaticPassage}, similar to \cite{mitra2020robust}. The adiabatic ramps are obtained by numerically maximizing the fidelity defined using the Hilbert-Schmidt overlap, 
\begin{equation}
    \mathcal{F}\left[\{c_{\mathrm{r}}\}\right] 
    = \frac{1}{16}
    \left \lvert \tr \left(
    \hat{U}_{\mathrm{MS}_{yy}}^{\dagger}
    \hat{U}(\{c_{\mathrm{r}}\})
    \right) \right\rvert^2,
\end{equation}
with respect to ramp parameters $\{c_{\mathrm{r}}\}$ for both one photon and two photon cases; here $\hat{U}(\{c_{\mathrm{r}}\})$ is the unitary map implemented using the spin-echo sequence in \cref{fig:SpinEchoAdiabaticPassage}. Replacing $\hat{H}$ with $\hat{H}_{\mathrm{eff}}$ gives an estimate of the fidelity including effects of finite lifetimes of the intermediate state $\ket{a}$ and the Rydberg state $\ket{r}$.

\begin{figure*}
    \includegraphics[width=0.92\textwidth]{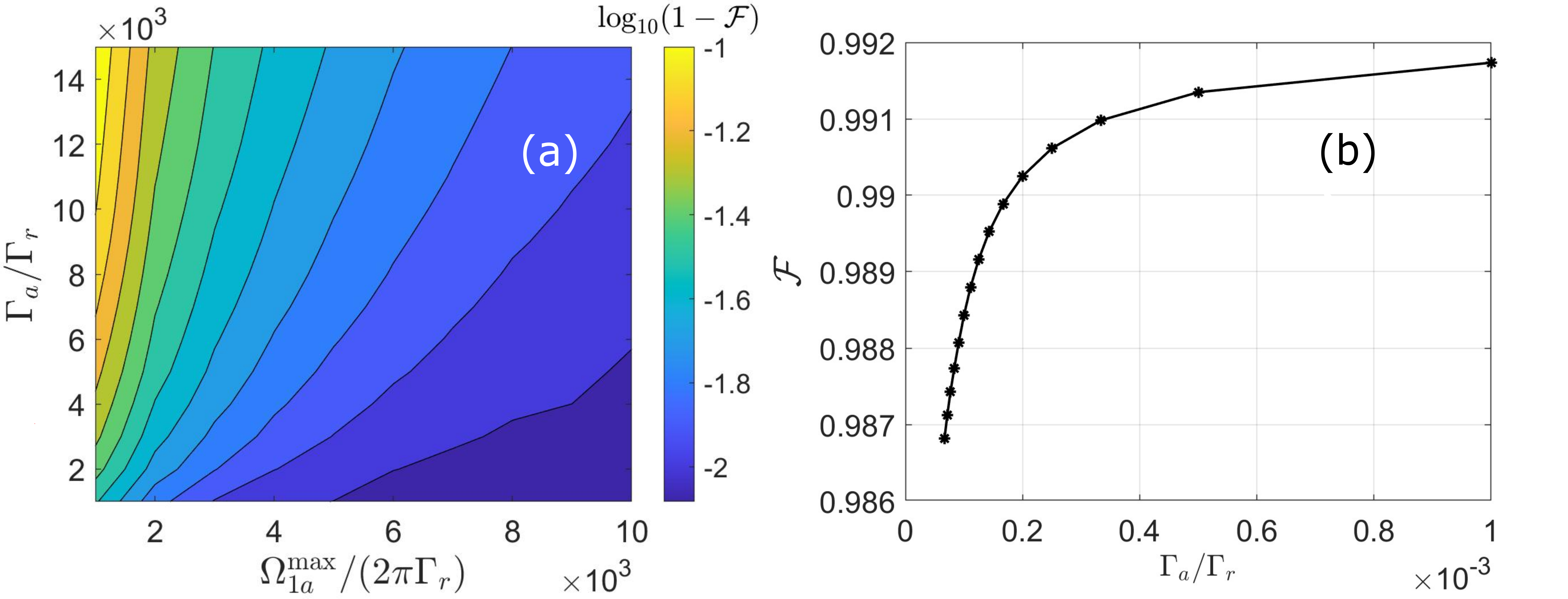}
    \caption{Dependence of the fidelity of the M{\o}lmer-S{\o}rensen gate on the intermediate state decay rate, $\Gamma_{a}$ and the Rabi frequency $\Omega_{1a}$, both measured in units of the Rydberg state decay rate $\Gamma_{r}$. Similar to other two photon approaches the choice of intermediate state with smaller decay rate gives a better fidelity. Moreover, as expected a larger power gives better fidelity. However, this gives us the constraint that we need a larger $|V|$ and thus posing some additional challenges. With a reasonable experimental parameters one could achieve an infidelity less than $10^{-2}$. The data are obtained by fixing the ratios $\Gamma_{a}/\Gamma_{r}$ and $\Omega_{1a}^{\max}/(2\pi\Gamma_{r})$ and optimizing over the choice of the detuning from the intermediate state $\Delta_{1a} = -\Delta_{ar}$. 
    (a) Contour plot of the logarithm of infidelity, $\log_{10}(1-\mathcal{F})$ across different values of $\Gamma_{a}/\Gamma_{r}$ and $\Omega_{1a}^{\max}/(2\pi\Gamma_{r})$.
    (b) Fidelity, $\mathcal{F}$ as a function of the ratio $\Gamma_{a}/\Gamma_{r}$, for $\Omega_{1r} = 1.4\Omega_{1a}^{\mathrm{max}}$ and $\Omega_{1a}^{\mathrm{max}}/(2\pi\Gamma_{r})= 10^{4}$.}
    \label{fig:FidelityPlots}
\end{figure*}

The short lifetime of the intermediate state $\ket{a}$ poses a challenge for implementing  adiabatic passage using a two-photon schemes.
We explore the dependence of the achievable M{\o}lmer-S{\o}rensen gate fidelity on the intermediate state lifetime and the Rabi frequency in \cref{fig:FidelityPlots}. We fix the Rydberg state decay rate $\Gamma_r$, vary the maximum Rabi frequency $\Omega_{1a}^{\max}$ and the intermediate state decay rate $\Gamma_a$, and then optimize over the intermediate state detuning $\Delta_{1a} = -\Delta_{ar}$ to maximize the fidelity. As in other two-photon approaches, the choice of an intermediate state with a larger lifetime gives a higher fidelity as this is the fundamental source of error in the model. Moreover, as expected a larger power gives higher fidelity, but in the perfect blockade regime this is constrained by $\hbar \Omega_{\mathrm{eff}} \ll \vert V \vert$. With reasonable experimental parameters, one can achieve fidelity larger than $0.99$  as seen in Fig. \ref{fig:FidelityPlots}.

A key metric quantifying the temporal duration of the adiabatic Rydberg dressing passages is the time-integrated Rydberg population, summed over both atoms, $t_r$  \cite{saffman2010quantum, saffman2016quantum}.
In order for the loss of fidelity due to Rydberg state decay to be small, we require $t_r \ll \tau_r$ where $\tau_r=1/\Gamma_r$ is the Rydberg state lifetime  \cite{mitra2020robust}.
For one photon adiabatic passages, we found $t_r \approx 0.89 \times 2\pi / \Omega_{\mathrm{eff}}^{\max}$, while for the two photon passage, we find $t_r \approx 0.95 \times 2\pi / \Omega_{\mathrm{eff}}^{\max}$, with initial state $\ket{1, 1}$. Initial states $\ket{0, 1}$ and $\ket{1, 0}$ lead to smaller time-integrated Rydberg population and initial $\ket{0, 0}$ does not lead to any Rydberg population \cite{mitra2020robust}. 
In both one and two-photon cases, since we are considering the strong blockade regime, the adiabatic passages, $t_r$ is significantly larger than $2\pi\hbar/|V|$, the time scale set by the interaction energy $V$.
Nevertheless, using finely tuned parameters, adiabatic Rydberg dressing passages can be used to implement high fidelity entangling gates.


\section{Dressing Beyond the Perfect Blockade Regime}
\label{sec:DressingEntanglingEnergy}
\begin{figure}
    \centering
    \includegraphics[width=0.48\textwidth]{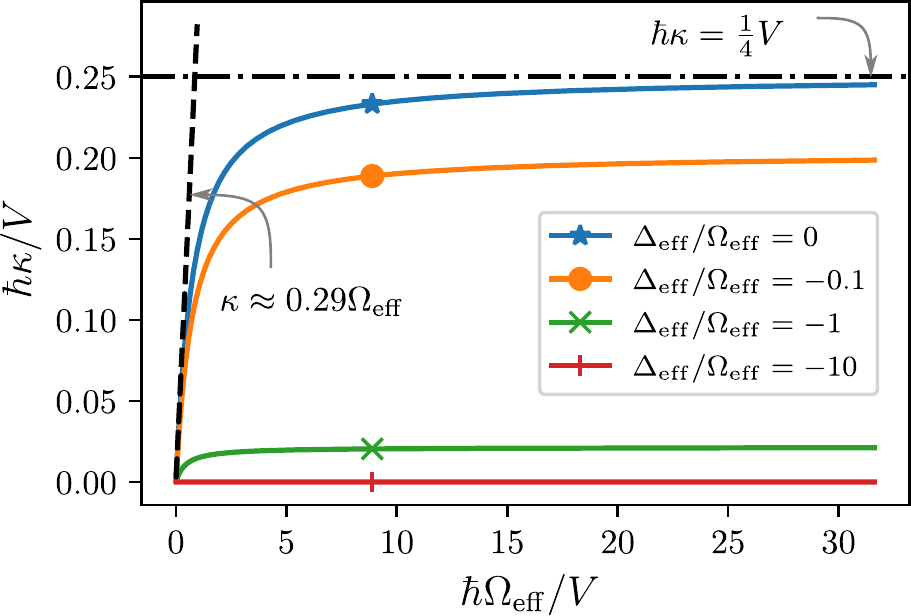}
    \caption{Entangling energy in units of the interaction energy as a function of the ground to Rydberg Rabi frequency in units of the interaction energy, for different detunings. For small detunings, in the strong blockade regime $\hbar\Omega_{\mathrm{eff}} \ll \vert V \vert$, the entangling energy scales linearly the Rabi frequency. and in the weak blockade regime $\hbar \Omega_{\mathrm{eff}} \gg \vert V \vert$, the entangling energy is independent of the Rabi frequency and scales linearly with the interaction energy. For large detunings, the entangling energy is negligible.}
    \label{fig:kappaOmegaVdd}
\end{figure}
\begin{figure*}
    \centering
    \includegraphics[width=0.9\textwidth]{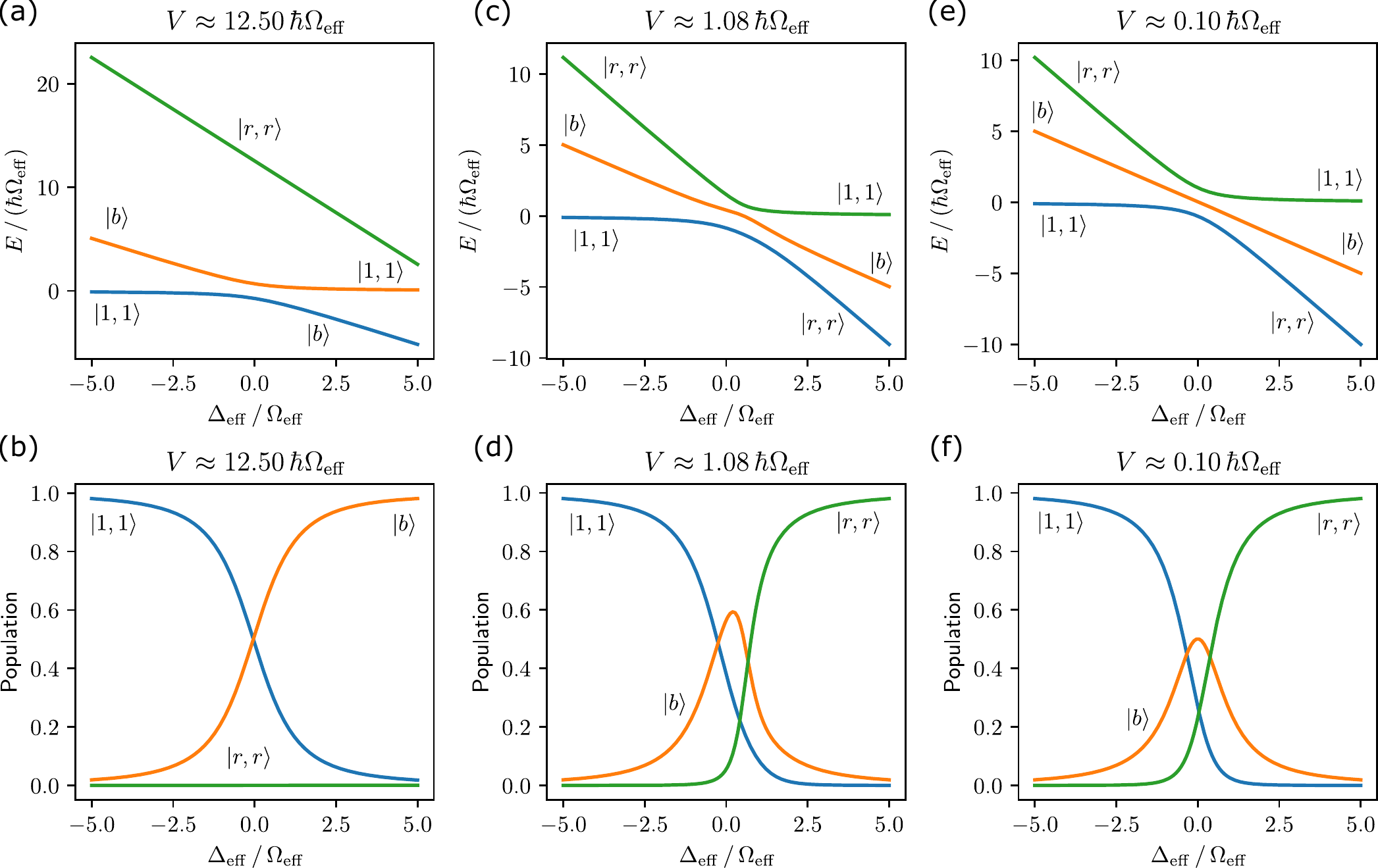}
    \caption{Dressed state energies and populations in the basis $\{\ket{1, 1}, \ket{b}, \ket{r, r}\}$ as a function of $\Delta_{\mathrm{eff}}/\Omega_{\mathrm{eff}}$ in different blockade regimes. (a, b) Strong blockade. (c, d) Intermediate blockade. (e, f) Weak blockade. (a, c, e) energy eigenvalues $V$, while (b, d, f) show populations of dressed states, when the initial detuning $\Delta_{\mathrm{eff}} < 0$.}
    \label{fig:BlockadeRegimesDressedStatePropertiess}
\end{figure*}
\begin{figure*}
    \centering
    \includegraphics[width=0.99\textwidth]{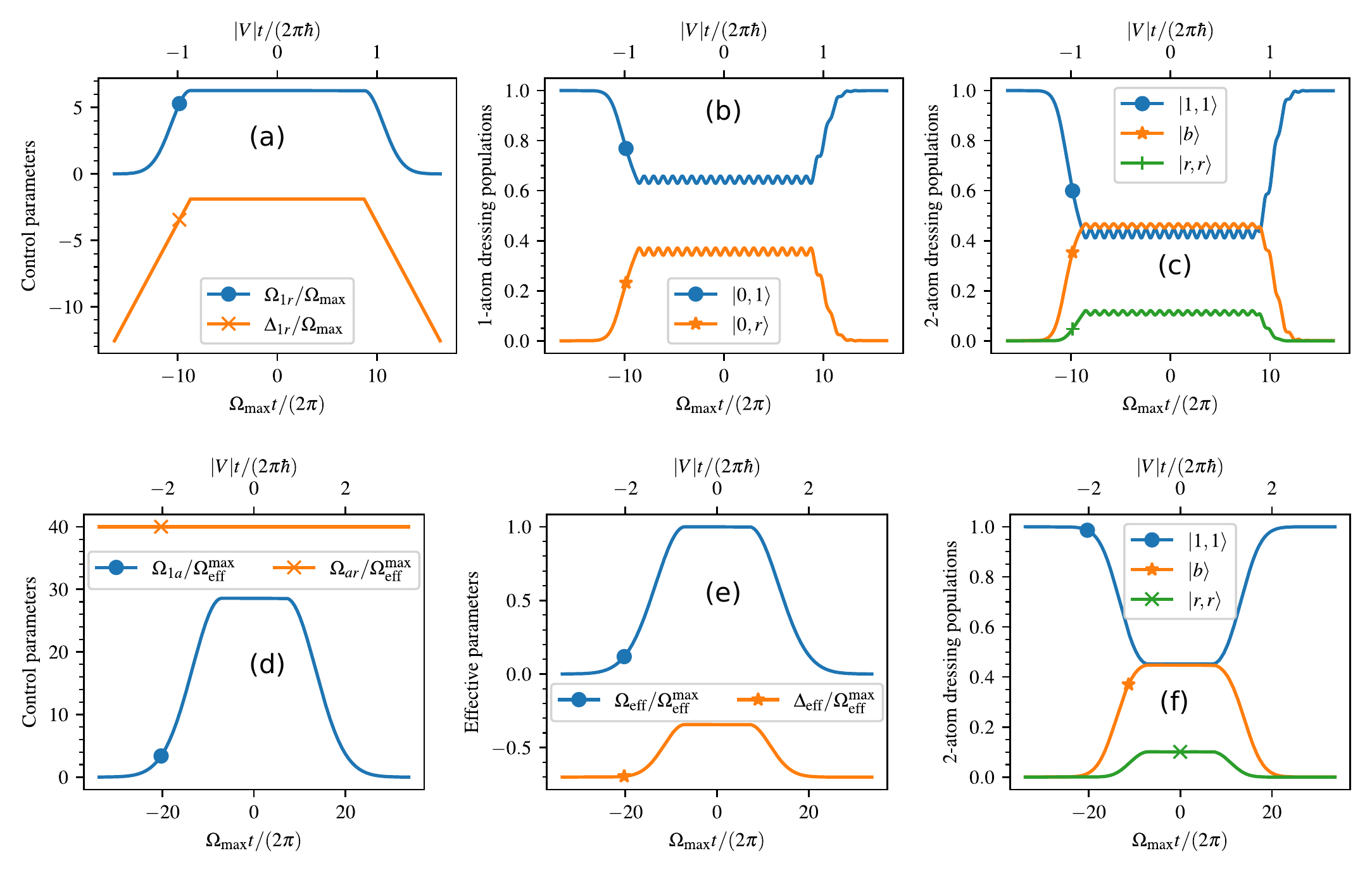}
    \caption{
    Adiabatic passages to implement $\hat{U}_{\kappa}(\varphi_1, \varphi_2)$ with $\varphi_2 = \pi/2$ [\cref{eq:UnitaryAdiabaticPassage}, \cref{fig:SpinEchoAdiabaticPassage}] in the weak blockade regime ($\hbar \Omega_{\mathrm{eff}} = 0.1 \vert V \vert$). (a) One-photon adiabatic passage Gaussian sweep of Rabi frequency and linear sweep of detuning as in \cite{mitra2020robust, martin2021molmer}. (b) 1-atom populations during a one-photon adiabatic passage  (c) 2-atom populations during a one-photon adiabatic passage. (d) Two-photon adiabatic passage using a Gaussian sweep of Rabi frequency $\Omega_{1a}$, with all other parameters fixed, which leads to an effective sweep of the two-photon Rabi frequency $\Omega_{\text{eff}}$ and two-photon detuning $\Delta_{\text{eff}}$ as shown in (e). (f) 2-atom populations during a two-photon adiabatic passage.
    Bottom axes show time measured in units of $2\pi/\Omega_{\max}$, top axes show time measured in units of $Vt/(2 \pi \hbar)$. In the weak blockade, as expected, $Vt/\hbar \ll \Omega_{\max}t$}
    
    \label{fig:WeakBlockadeAdiabaticPassage}
\end{figure*}
\begin{figure}
    \includegraphics[width=0.48\textwidth]{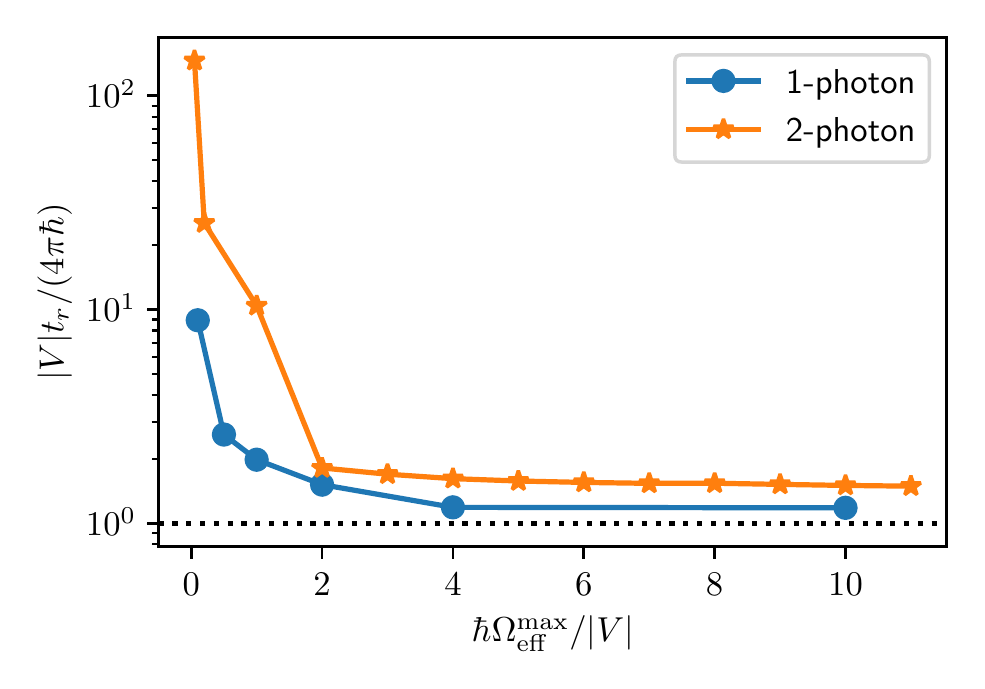}
    \caption{Time integrated Rydberg population $t_r$ as a function of Rydberg interaction $V$ for the one- and two- photon ramps. In both cases, the integrated Rydberg population becomes lower and lower as we increase $\hbar \Omega_{\mathrm{eff}} / |V|$ for the two-photon adiabatic passage as in \cref{eq:GaussianOmega_a} and  One-photon adiabatic passage as in \cref{eq:GaussianOmega_1rLinearDelta_1r}
    }
    \label{fig:IntegratedRydbergPopulation}
\end{figure}
 In the previous section, we studied entangling gates in the case of a perfect Rydberg blockade, but this is not intrinsic to the adiabatic dressing protocol. 
 Relaxing this assumption and studying protocols in the weak blockade regime is important in order to address the fundamental limits of Rydberg-atom quantum information processing, potentially improve the fidelity of our gates, and allow us to operate in new regimes.
 Implementation of an entangling gate using Rydberg-meditated interactions is fundamentally limited by two energy-time scales -- the Rydberg state lifetime $\tau_r$ and the magnitude of the interatomic interaction energy $|V|$  \cite{saffman2010quantum, saffman2016quantum}. 
 Wesenberg {\em et al.} showed that the minimum time that the atoms need to spend in a Rydberg state to achieve a maximally entangling gate scales as $t_{r} \sim \hbar/|V| $ \cite{wesenberg2007scalable}. 
 The standard protocols which employ a strong Rydberg blockade \cite{jaksch2000fast, levine2019parallel} cannot achieve this bound because the speed of the gates is set by $\Omega_{\rm{eff}}$, and since they require $\hbar \Omega_{\rm{eff}}\ll \vert V \vert$, we cannot make use of the full scale of the interaction energy \cite{saffman2010quantum}. Jo {\em et. al.} implemented Rydberg mediated entanglement outside the strong blockade regime using finely tuned two-atom Rabi oscillations \cite{jo2020rydberg}.

The fundamental scaling can be understood in a simple protocol using the limiting case of very large Rabi frequency $\hbar \Omega_{\mathrm{eff}}^{\max} / |V| \to \infty$. 
An entangling gate can be achieved using a collective $\pi$-pulse from $\ket{1}$ to $\ket{r}$ on both atoms, followed by an interaction for a time $|V|t/\hbar = \pi$ and a $\pi$-pulse from $\ket{r}$ to $\ket{1}$. In the limit of infinitesimally short $\pi$-pulses, the time spent in Rydberg states, or time-integrated Rydberg population, is $\pi\hbar/|V|$.
All of the time spent in the Rydberg states is in the doubly excited Rydberg state $\ket{r,r}$.

 While this simple protocol helps us understand the fundamental scaling, it is generally not practical for implementation. 
 For small interatomic separations, the two-atom spectrum becomes a complex tangle of ``Rydberg spaghetti" \cite{keating2013adiabatic, jau2016entangling}.
 To achieve the fastest gates in this strongly interacting case, it is thus useful to avoid double Rydberg population which can lead to unexpected inelastic processes.
 In addition the complex potential landscape at such small interatomic separations can lead to high sensitivity to atomic motion. 
 In this section we show that using adiabatic Rydberg dressing, we can get close to the fundamental bound, while working in the weak blockade regime, $\hbar \Omega_{\mathrm{eff}}^{\max} \gg |V|$, without significant double Rydberg population.
 Moreover, for large interatomic separations protocols requiring a strong blockade would lead to exceedingly slow gates. 
 The adiabatic dressing protocol considered here can achieve reasonably fast gates with high fidelity even for atoms separated beyond blockade radius.

To understand the different regimes of operation, we estimate how the interatomic interaction energy $V$ limits the entangling energy $\hbar \kappa$ in the strong blockade and weak blockade regimes. 
For simplicity, we consider the case in which the atoms see the same Rabi frequency, given in \cref{eq:TwoAtomRydbergHamiltonian}. 
 It is useful to consider a pseudo-spin with $\ket{\uparrow_z} \equiv \ket{r}$ and $\ket{\downarrow_z} \equiv \ket{1}$. 
 Note that this is different from the dressed pseudo-spin considered in  \cite{mitra2020robust, martin2021molmer, schine2022long}, where the pseudo-spin levels corresponded to the dressed ground states. In this pseudo-spin picture, the two-atom Hamiltonian can be written as a sum of two terms
 \begin{equation}
 \begin{aligned}
    \hat{H}_{\text{int}}
	& = V \ketbra{r,r}{r,r} \equiv \frac{V}{2} \qty(\hat{S}_z^2 + \hat{S}_z),
	\\
	\hat{H}_{\text{drive}} 
	& \equiv -\hbar\Delta_{\mathrm{eff}} \, \mathbbm{1} - \hbar\Delta_{\mathrm{eff}} \, \hat{S}_z + \hbar\Omega_{\mathrm{eff}} \, \hat{S}_x
	\\ & 
	\equiv -\hbar\Delta_{\mathrm{eff}} \, \mathbbm{1} + \hbar\sqrt{\Delta_{\mathrm{eff}}^2 + \Omega_{\mathrm{eff}}^2} \, \hat{S}_\theta,     
 \end{aligned}
 \end{equation}
where $\hat{S}_{\mu}$ is the $\mu$-component of collective angular momentum operator $S_\mu = \mathbbm{1} \otimes \hat{\sigma}_\mu/2 + \hat{\sigma}_\mu/2 \otimes \mathbbm{1}$, $\hat{S}_\theta = \cos\theta \hat{S}_z + \sin\theta \hat{S}_x$ with $\tan \theta = \Omega_{\mathrm{eff}}/(-\Delta_{\mathrm{eff}})$. The collective symmetric spin-1 eigenstates of $S_z$ are the triplet of the pseudospins $\ket{S=1, M_z=-1}=\ket{1,1}$, $\ket{S=1, M_z=0}=(\ket{1,r}+\ket{r,1})/\sqrt{2} =\ket{b}$, $\ket{S=1, M_z=+1}=\ket{r,r}$. The eigenvalues and eigenvectors of the driving Hamiltonian and the interaction Hamiltonian are in \cref{tab:EigenAtomLight} and \cref{tab:EigenAtomAtom} respectively. 

\begin{table}[]
    \centering
    \begin{tabular}{ |c|c| } 
 \hline
 $\text{Energy Eigenvalue}$ &$\text{Eigenvectors}$  \\ 
\hline
$-\hbar \Delta_{\mathrm{eff}} +\hbar \sqrt{\Omega_{\mathrm{eff}}^2 + \Delta_{\mathrm{eff}}^2}$ & 
$\ket{\uparrow_\theta} \otimes \ket{\uparrow_\theta}$ \\
 \hline
 $-\hbar \Delta_{\mathrm{eff}} -\hbar \sqrt{\Omega_{\mathrm{eff}}^2 + \Delta_{\mathrm{eff}}^2}$ & 
$\ket{\downarrow_\theta} \otimes \ket{\downarrow_\theta}$ \\
\hline
  $-\hbar \Delta_{\mathrm{eff}}$ & 
  $\left(\ket{\uparrow_\theta} \otimes \ket{\downarrow_\theta} + \ket{\downarrow_\theta} \otimes \ket{\uparrow_\theta}\right)/\sqrt{2}$\\
\hline
\end{tabular}
    \caption{Eigenvalues and eigenvectors of the atom-light Hamiltonian, $\hat{H}_{\mathrm{drive}}$. Here $\ket{\uparrow_\theta} \equiv \cos\qty(\theta/2) \ket{\uparrow_z} + \sin\qty(\theta/2) \ket{\downarrow_z}$, $\ket{\downarrow_\theta} \equiv \cos\qty(\theta/2) \ket{\downarrow_z} - \sin\qty(\theta/2) \ket{\uparrow_z}$ and $\tan\theta = \Omega_{\mathrm{eff}} / (-\Delta_{\mathrm{eff}})$. The first two rows represent the upper and lower branches of the single atom dressed states.}
\label{tab:EigenAtomLight}
\end{table}


\begin{table}[]
    \centering
    \begin{tabular}{ |c|c|c| } 
 \hline
 $\text{Energy Eigenvalue}$ &$\text{Eigenvectors}$  \\ 
\hline
 $V$  & $\ket{r,r}$ \\
\hline
     $0$  &
     $\ket{b},\;\ket{1,1}$ \\
\hline
\end{tabular}
\caption{Eigenvalues and eigenvectors of the atom-atom interaction Hamiltonian, $\hat{H}_{\mathrm{int}}$, in the symmetric subspace, spanned by $|1,1\rangle, |b\rangle, |r,r\rangle$.}
\label{tab:EigenAtomAtom}
\end{table}

First we consider the well-known strong blockade regime with $\vert V \vert \gg \hbar \Omega_{\mathrm{eff}}$, where the interaction term is the dominant Hamiltonian and the driving term is the perturbation.  The zeroth order eigenvectors are the states $\ket{S=1,M_z}$. The leading order correction is calculated using degenerate perturbation theory in the zero eigenvalue subspace spanned by $\ket{S=1, M_z=-1} \equiv \ket{1,1}$ and $\ket{S=1, M_z=0} \equiv \ket{b}$. Using $\mathscr{P}_{S, M_z}$ to denote the projector on the subspace of $S, M_z$,
\begin{equation}
\begin{aligned}
    &
    \qty(\mathscr{P}_{S=1, M_z=-1} + \mathscr{P}_{S=1, M_z=0})
    \hat{S}_{\theta}
    \\ &
    \qty(\mathscr{P}_{S=1, M_z=-1} + \mathscr{P}_{S=1, M_z=0})
    \\ &
    = - \cos(\theta) \ketbra{S=1, M_z=-1}{S=1, M_z=-1}
    \\ &
    + \frac{\sin(\theta)}{\sqrt{2}} \qty(
    \ketbra{S=1, M_z=-1}{S=1, M_z=0}) 
    \\ &
    + \frac{\sin(\theta)}{\sqrt{2}} \qty(
    \ketbra{S=1, M_z=0}{S=1, M_z=-1}).
\end{aligned}
\end{equation}

The perturbative corrections to energy eigenvalues are the two-atom light shift experienced by the atoms together, in the presence of $V$.
The leading correction to the energy of the logical state $\ket{1, 1} \equiv \ket{S=1,M_z=-1}$ in perturbation theory, is the two-atom light shift under perfect blockade,
 \begin{equation}
     E_{\text{LS}}^{(2)} = -\frac{\hbar \Delta_{\mathrm{eff}}}{2} \pm \frac{\hbar}{2} \sqrt{2\Omega_{\mathrm{eff}}^2 + \Delta_{\mathrm{eff}}^2}.
 \end{equation}
Subtracting out the energy shifts in eigenstates of each atom to obtain the entangling energy $\kappa$ using \cref{eq:EntanglingEnergy},
\begin{equation}
\begin{aligned}
    &
    \lim_{\hbar\Omega_{\mathrm{eff}} / \vert V \vert \to 0}
    \kappa 
    \\ & 
    = - \frac{\Delta_{\mathrm{eff}}}{2} 
    \pm 
    \frac{1}{2}\qty(\sqrt{\Delta_{\mathrm{eff}}^2 + 2\Omega_{\mathrm{eff}}^2}
    - 2 \sqrt{\Delta_{\mathrm{eff}}^2 + \Omega_{\mathrm{eff}}^2}).
\end{aligned}
\end{equation}
 Note that here by design, $\hbar \vert \kappa \vert \ll \vert V \vert$, since we assumed $\hbar\Omega_{\mathrm{eff}} \ll \vert V \vert$. The maximum useful $\kappa$ scales with the Rabi frequency $\Omega_{\mathrm{eff}}$. Under a perfect Rydberg blockade regime $\vert V \vert \gg \hbar \Omega_{\mathrm{eff}}$, the state $\ket{r,r}$ is not populated. Thus, there is an adiabatic passage from the $\ket{1, 1}$ to $\ket{b}$ and back as shown in \cref{fig:EnergyLevels}(f).

\begin{table}[]
    \centering
    \begin{tabular}{ |c|c| } 
 \hline
 $\text{Energy Eigenvalue}$ & $\text{Eigenvectors}$  \\ 
 \hline
     $-\frac{1}{2} \cos\theta + \frac{1}{2} \sqrt{\cos^2\theta + 2\sin^2\theta}$  &
     $\cos\frac{\Theta}{2} \ket{b} + \sin\frac{\Theta}{2} \ket{1,1}$
\\
\hline $-\frac{1}{2} \cos\theta - \frac{1}{2} \sqrt{\cos^2\theta + 2\sin^2\theta}$  &
     $\cos\frac{\Theta}{2} \ket{1,1} - \sin\frac{\Theta}{2} \ket{b}$
      \\
\hline
\end{tabular}
    \caption{Eigenvalues and eigenvectors of $\hat{S}_{\theta}$ in the zero-eigenvalue subspace of $\hat{H}_{\mathrm{int}}$. Here, $\tan \Theta = \sqrt{2}\Omega_\mathrm{eff} / (-\Delta_{\mathrm{eff}})$. The upper and lower rows represented the upper and lower branches of the two-atom dressed states in the perfect blockade regime, shown in \cref{fig:EnergyLevels}.}
\label{tab:EigenSthetaInProjAtomAtom}
\end{table}

Next, we consider the weak blockade regime where $\vert V \vert \ll \hbar \Omega_{\mathrm{eff}}$. In this case, the laser driving term is the dominant Hamiltonian and the interaction term is a perturbation. The eigenstates of the driving Hamiltonian are the one-atom dressed states, which are rotated spin-triplet states $\ket{S=1, M_\theta}$ given in \cref{tab:EigenAtomLight}. The energy eigenvalues are the one-atom light shift. The entangling energy $\hbar\kappa$ can be estimated as the correction to the dress-ground state $\ket*{\widetilde{1,1}} \equiv \ket{S=1, M_\theta=-1} \equiv \left( \cos\frac{\theta}{2} \ket{1} + \sin\frac{\theta}{2}\ket{r}\right)^{\otimes 2}$. The unperturbed energies of the dominant Hamiltonian include the single-atom light shifts. Therefore the leading order correction to the non-interacting energy is the asymptotic value of $\hbar \kappa$,
\begin{equation}
\begin{aligned}
    &
    \lim_{ \vert V \vert / \hbar\Omega_{\mathrm{eff}} \to 0}
	\hbar \kappa 
	= \qty(\frac{1 \pm \cos \theta}{2})^2 V,
	\label{eq:kappaAsymptoteWeakBlockade}
\end{aligned}
\end{equation}
where $\pm$ refers to the relative sign of the initial detuning and the detuning at peak dressing during an adiabatic passage, and the corresponding (unnormalized) dressed state, in leading order perturbation theory, is
\begin{equation}
\begin{aligned}
    \ket*{\widetilde{1, 1}}
    &
    \equiv
    \qty(\cos\frac{\theta}{2}\ket{1} + 
    \sin\frac{\theta}{2}\ket{r})^{\otimes 2} 
    \\ &
    \pm
    \cos^2\left(\frac{\theta}{2}\right)\;
    \frac{V}{2\hbar \sqrt{\Omega_{\mathrm{eff}}^2 + \Delta_{\mathrm{eff}}^2}}
    \ket{r, r},
\end{aligned}
\end{equation}
now including the doubly excited Rydberg state.

We calculate the entangling energy $\hbar\kappa$ numerically beyond the perfect blockade regime for different detunings as shown in~\cref{fig:kappaOmegaVdd}. We focus on entangling protocols that limit the population in the doubly-excited Rydberg state, $\ket{r, r}$, in order to avoid potentially deleterious decay and inelastic processes. To ensure this, we consider adiabatic ramps that are far from the anti-blockade condition, $V = 2\hbar\Delta_{\mathrm{eff}}$. In practice, this is done in the weak blockade case with a detuning at peak dressing (minimum $|\Delta_{\mathrm{eff}}|$) satisfying $\hbar |\Delta_{\mathrm{eff}}| \ll |V|$. As predicted from perturbation theory, we see that entangling energy scales with the Rabi frequency in the strong blockade regime and reaches $V/4$ at resonance, in the weak blockade regime.

Theoretically, all of the interaction energy $V$ is available as the Rydberg dressing entangling energy $\hbar \kappa$. However, this occurs when $\theta \in \qty{0, \pi}$ or $\vert \Delta_{\mathrm{eff}} \vert / \Omega_{\mathrm{eff}} \to \infty$ when the dressed state is simply the bare atomic state $\ket{r, r}$. As we saw in \cite{mitra2020robust}, an adiabatic passage that starts far from ground-Rydberg resonance, goes close to resonance, and returns to far off-resonance is most effective at limiting double Rydberg excitation. In this weak blockade case the adiabatic passage stays far from the anti-blockade condition, leading to a dressed state $\ket*{\widetilde{1, 1}}$ that is primarily an admixture of $\ket{1, 1}$ and the bright state $\ket{b}$, with a small $\ket{r, r}$ component.

In \cref{fig:BlockadeRegimesDressedStatePropertiess} we consider examples of strong ($V\gg \hbar\Omega_{\mathrm{eff}}$), intermediate ($V\sim \hbar\Omega_{\mathrm{eff}}$), and weak ($V \ll \hbar\Omega_{\mathrm{eff}}$), showing the dressing energies and the populations of bare states $\ket{1, 1}, \ket{b}, \ket{r, r}$ in the dressed state $\ket*{\widetilde{1, 1}}$. Given the energy gaps, we see that the adiabatic dressing protocol allows for a gate as fast as a time scale of $\sim 2\pi\hbar/|V|$, and importantly, by sweeping the detuning close to resonance, while avoiding the anti-blockade condition, there is negligible excitation of the doubly excited Rydberg state $\ket{r, r}$. For example, we study $|V| = 0.1\hbar\Omega_{\mathrm{eff}}$ for both the one and two-photon excitation; the ramps are shown in \cref{fig:WeakBlockadeAdiabaticPassage} using the same parameterization used for the strong blockade case, \cref{eq:GaussianOmega_1rLinearDelta_1r}, and \cref{eq:GaussianOmega_a}. Despite the weak blockade, we see that the population accumulated in the state $\ket{r,r}$ is bounded, which overcomes one of the significant hurdles in going beyond perfect blockade.

Let us return to the question of the maximum possible achievable entangling gate fidelity. When considering adiabatic Rydberg dressing, the entanglement is generated in the form of the dynamical phases from the entangling energy, $\int \dd t' \kappa(t')$  \cite{keating2015robust, mitra2020robust}. Fundamentally, the time spent in the Rydberg state is bounded by an energy scale proportional to the entangling energy $\hbar \kappa$. Using adiabatic Rydberg dressing in the strong-blockade regime leads to $t_{r}$ that scales inversely with the Rabi frequency as $\kappa \sim \Omega_{\max}$, and therefore is far from the minimum, $t_r \sim 2\pi/\Omega_{\max} \gg \pi \hbar/\vert V \vert$. In \cref{fig:IntegratedRydbergPopulation}, we plot the time-integrated Rydberg population as a function of  the ratio of Rabi frequency to the interatomic Rydberg interaction energy $\hbar \Omega_{\mathrm{eff}}^{\mathrm{max}}/ \vert V \vert$ for both the one-photon case using \cref{eq:GaussianOmega_1rLinearDelta_1r} and the two-photon ramps as given in \cref{eq:GaussianOmega_a}. The analysis indicates that the time-integrated population required to create the perfect entangler, while avoiding the anti-blockade condition, decreases as we increase the Rabi frequency $\hbar\Omega_{\max}$, compared to the interaction energy $|V|$ and it eventually saturates to slightly above $4\pi \hbar/|V|$.

This result is consistent with the bound found in \cite{wesenberg2007scalable}. 
Since the value of $\hbar |\kappa|$ reaches $|V|/4$, near resonance in the weak blockade regime [\cref{eq:kappaAsymptoteWeakBlockade}], the theoretically achievable maximum fidelity, while limiting double Rydberg excitation is
\begin{equation}
     \mathcal{F} < 1 - \frac{4\pi\hbar}{\vert V \vert \tau_r},
\end{equation}
where $\tau_r$ is the Rydberg state lifetime. For contemporary experiments, with $\vert V \vert /(2\pi\hbar) = 40\;\mathrm{MHz}$ and $\tau_r = 150 \mathrm{\mu s}$, the theoretical minimum infidelity is about $10^{-3}$. With cryogenically enhanced Rydberg lifetimes, around $\tau_r = 1\mathrm{ms}$ and stronger interactions, $\vert V \vert/(2\pi\hbar) = 1$ GHz, the theoretical minimum infidelity would be $10^{-5}$. In practice achieving this would require working in the weak blockade regime, with large laser power such that $\hbar\Omega_{\text{eff}} \gg \vert V \vert$.

The ability to design gates with adiabatic dressing beyond the perfect blockade regime also loosens other constraints and potential sources of error. Maintaining atoms beyond the blockade radius reduces the requirement for transporting atoms, which leads to motional heating. Our results show that even for moderate EDDI, with $|V|/(2\pi\hbar)$ of a few $\mathrm{MHz}$, one can achieve fast gates with gates times of the order of a few $\mu$s. Moreover, at moderate separations the shifted doubly excited states are well resolved and well defined, reducing spurious resonances. A potential downside to operation in this regime is the sensitivity of the entangling energy to atom separation and also the resulting forces on the atoms. We address this in \cref{sec:Forces}.


\section{Conclusion}
\label{sec:Conclusions}

In this article, we explored some practical considerations and the fundamental limits of the adiabatic Rydberg dressing protocol for two-qubit quantum logic gates, where entanglement is generated by the modification of ground state light shift introduced by the interaction energy between Rydberg atoms. We studied adiabatic Rydberg dressing via a two-photon ground-to-Rydberg transition and found adiabatic ramps that can be used to achieve high fidelity entangling gates by modulating only one laser amplitude as a function of time, with all laser frequencies fixed, allowing an easier experimental implementation and alleviating the need for a high power ultraviolet laser (\cref{sec:TwoPhotonAdiabaticPassage}). A major bottleneck for adiabatic Rydberg dressing-based entangling gates in the case of a two-photon ground-to-Rydberg transition is the intermediate state lifetime. We found that with the current state of the art, gates with fidelity $>0.99$ are achievable in a regime that adiabatically eliminates the intermediate state, but still maintains reasonable two-photon Rabi frequencies. This protocol is applicable for both alkali and alkaline-earth like atoms.

We also studied the fundamental limits of implementing an entangling gate using adiabatic Rydberg dressing of ground states, which are based on the finite Rydberg lifetime and the entangling energy obtained in the dressed states (\cref{sec:DressingEntanglingEnergy}). We showed that in the well-known strong blockade regime, the entangling energy scale is limited by the ground-Rydberg Rabi frequency, that is, laser power, and in the weak blockade regime, the entangling energy is limited by the interaction energy between the atoms. Moreover, we showed proof-of-principle feasibility of rapid adiabatic passages without significant double-Rydberg population in strong, intermediate, and weak blockade regimes, thereby loosening the requirements of atoms being within a blockade radius for implementing entangling gates in a few $\mathrm{\mu s}$. A more precise model of the entangling energy using atomic species and Rydberg state specific treatment, for example as in  \cite{de2018analysis}, can be used to design adiabatic passages for specific experiments.

In conclusion, adiabatic Rydberg dressing is a promising approach to implementing two-qubit entangling gates for neutral atoms. It can be implemented in several atomic species with one- or two-photon ground-to-Rydberg transitions and can be designed beyond the strong blockade regime to yield fast, high fidelity gates.

\begin{acknowledgements}
We thank Antoine Browaeys for the fruitful discussions.
This research is supported by the Laboratory Directed Research and Development program of Los Alamos National Laboratory under project number 20200015ER and the NSF Quantum Leap Challenge Institutes program, Award No. 2016244.
Founded in 1889, the University of New Mexico sits on the traditional homelands of the Pueblo of Sandia. The original peoples of New Mexico -- Pueblo, Navajo, and Apache -- since time immemorial, have deep connections to the land and have made significant contributions to the broader community statewide. We honor the land itself and those who remain stewards of this land throughout the generations and also acknowledge our committed relationship to Indigenous peoples. We gratefully recognize our history.
Los Alamos National Laboratory is managed by Triad National Security, LLC, for the National Nuclear Security Administration of the U.S. Department of Energy under Contract No. 89233218CNA0000. 
\end{acknowledgements}

\appendix
\section{Quantifying force on atoms}
\label{sec:Forces}

Outside the strong blockade regime, it is important to consider the interatomic forces that could potentially effect the motional state of the atoms. Two atoms directly excited into the Rydberg state will experience a large Van der Waals force from  the EDDI. However, in adiabatic dressing a force will arise from the spatial gradient of the light shift, i.e., the ``soft core" adiabatic potential force arising from the $\ket{r, r}$ component in the dressed state $\ket*{\widetilde{1,1}}$.

Consider, thus, the adiabatic  interatomic potential experienced by atoms in instantaneous internal ``adiabatic state" $\ket{\psi(\vb{R})}$. We treat here the center of mass motion of the atoms classically, in which the interatomic force is given according to
\begin{equation}
	V_{\text{ad}} 
	= \bra{\psi(\vb{R})} H(\vb{R}) \ket{\psi(\vb{R})}
	\implies \vb{F} 
	= -\grad V_{\text{ad}},
\end{equation}
where $\ket{\psi(\vb{R})} = c_{11}(\vb{R}) \ket{1,1} + c_b(\vb{R}) \ket{b} + c_{rr} \ket{r,r}$. The coefficients depend on the interatomic distance $\vb{R}$. If the state is an eigenstate of $H$, for example, the adiabatic potential of the dressed ground state $\ket*{\widetilde{1,1}}$ is
\begin{equation}
	V_{\text{ad}} (\widetilde{1, 1}) 
	= E(\widetilde{1, 1}) 
	= \hbar\kappa(\vb{R}) + 2 E_{\text{LS}}^{(1)} + 2 E_{1},
\end{equation}
which gives a force
\begin{equation}
	\vb{F}(\widetilde{1, 1}) = -\hbar \grad \kappa(\vb{R}),
\end{equation}
as the one-atom light shift $E_{\text{LS}}^{(1)}$ and bare energy $E_1$ are independent of $\vb{R}$.

When the interatomic distance is well within the blockade radius, where we have a perfect blockade, $\kappa$ is 
independent of $\vb{R}$. This leads to a ``soft-core'' potential which has been observed experimentally \cite{jau2016entangling, zeiher2016many, zeiher2017coherent, martin2021molmer}. This can also be analyzed on the basis of bare atomic orbitals. For simplicity consider the case of zero detuning, the adiabatic potential between dressed ground states is 
\begin{equation}
\begin{aligned}
 	 V_{\text{ad}} (\widetilde{1, 1})  &
	 = \sqrt{2} \Omega_{\mathrm{eff}} (
	 \Re \qty(c_{11}(\vb{R}) c_{b}^{*}(\vb{R}))
	 \\ & +
	 \Re \qty(c_{11}(\vb{R}) c_{rr}^{*}(\vb{R})) )
	 + \vert c_{rr}\vert^2 V(\vb{R}).
\end{aligned}
\end{equation}
Note, the force is not simply $\vert c_{rr}\vert^2 \grad V(\vb{R})$; the interference terms in the adiabatic potential reduces the otherwise large force.

\begin{figure}
    \centering
    \includegraphics[width=0.48\textwidth]{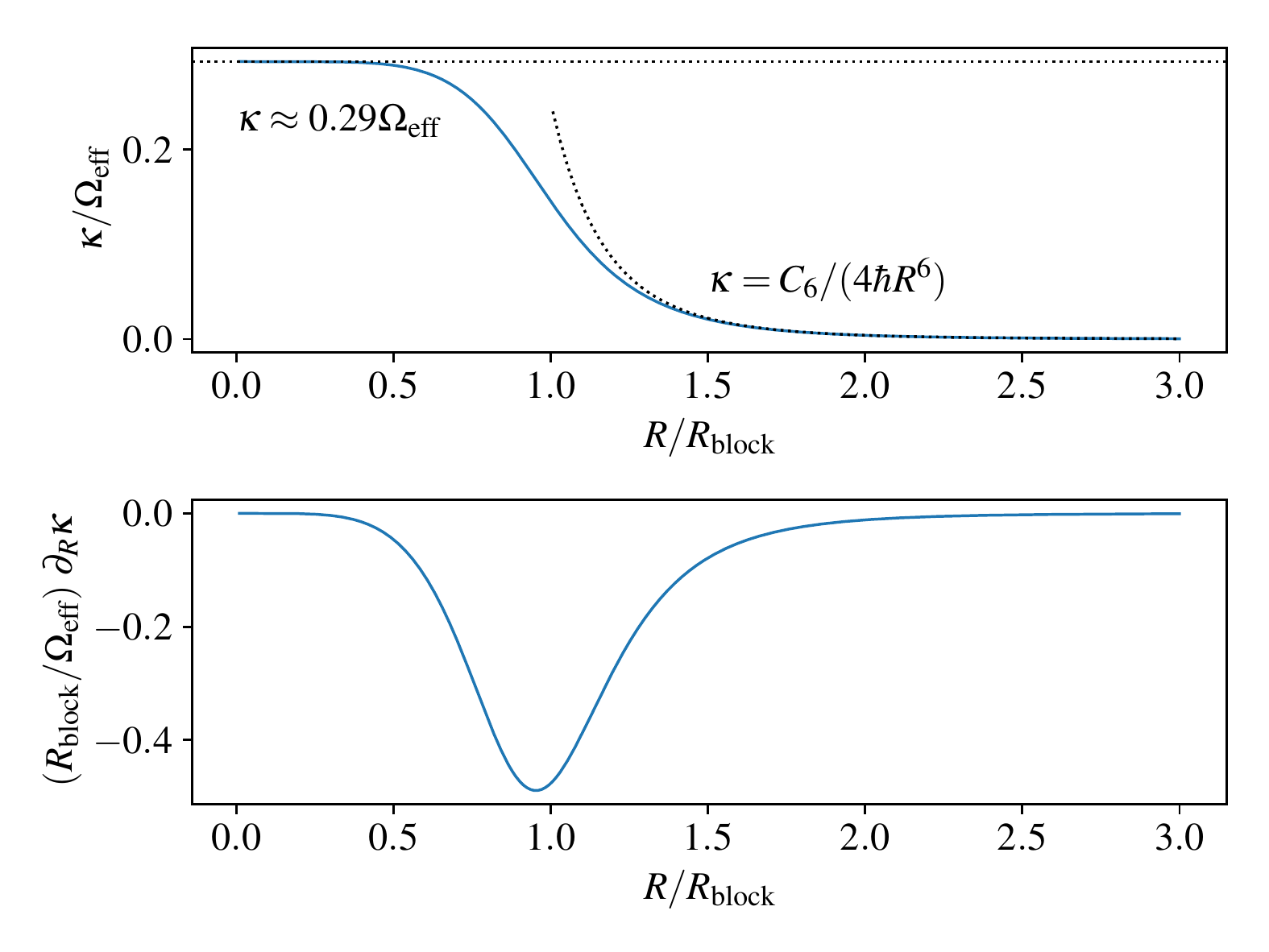}
    \caption{Interatomic potential energy and forces between atoms in the dressed state $\ket*{\widetilde{1,1}}$.
    (a) Entangling energy $\kappa$ in units of the Rabi frequency $\Omega_{\text{eff}}$ as a function of inter-atomic distance $R$, in units of the blockade radius $R_{\text{block}}$.
    (b) Gradient of the entangling energy along the inter-atomic direction $\partial_{R} \kappa$ in units of the ratio $\Omega_{\text{eff}}/R_{\text{block}}$.
    }
    \label{fig:InteratomicPotentialEnergy}
\end{figure}

For simplicity and generality, we calculate $\kappa$ as a function of distance using a Van der Waals potential, $V = C_6 |\vb{R}|^{-6}$,
and the interatomic force as a function of distance in \cref{fig:InteratomicPotentialEnergy}. As is standard, we define the blockade radius where the energy of Rabi frequency of Rydberg excitation is equal to $V$, $\hbar \Omega_{\mathrm{eff}} = C_6 R_{\text{block}}^{-6}$. At short interatomic distances the adiabatic potential has a soft-core form and is the entangling energy $\hbar \kappa$, up to additive constants, as observed experimentally in  \cite{jau2016entangling, zeiher2016many, zeiher2017coherent, martin2021molmer}. At large distances, the interatomic potential asymptotes to a quarter of the vans Der Waals potential, $C_6 \left|\vb{R}\right|^{-6}/4$ for Van der Waals interactions. The transition occurs roughly between $\left|\vb{R}\right| /R_{\text{block}}\approx 1/2$ and $\left|\vb{R}\right| /R_{\text{block}}\approx 2$, where the potential energy has a nonzero gradient, giving rise to a non-trivial interatomic force (\cref{fig:InteratomicPotentialEnergy}). From these results, we see that the operation of an adiabatic dressing gate outside the perfect blockade regime will lead to bounded perturbing forces on the atoms.

\bibliography{references}

\begin{thebibliography}{47}%
\makeatletter
\providecommand \@ifxundefined [1]{%
 \@ifx{#1\undefined}
}%
\providecommand \@ifnum [1]{%
 \ifnum #1\expandafter \@firstoftwo
 \else \expandafter \@secondoftwo
 \fi
}%
\providecommand \@ifx [1]{%
 \ifx #1\expandafter \@firstoftwo
 \else \expandafter \@secondoftwo
 \fi
}%
\providecommand \natexlab [1]{#1}%
\providecommand \enquote  [1]{``#1''}%
\providecommand \bibnamefont  [1]{#1}%
\providecommand \bibfnamefont [1]{#1}%
\providecommand \citenamefont [1]{#1}%
\providecommand \href@noop [0]{\@secondoftwo}%
\providecommand \href [0]{\begingroup \@sanitize@url \@href}%
\providecommand \@href[1]{\@@startlink{#1}\@@href}%
\providecommand \@@href[1]{\endgroup#1\@@endlink}%
\providecommand \@sanitize@url [0]{\catcode `\\12\catcode `\$12\catcode
  `\&12\catcode `\#12\catcode `\^12\catcode `\_12\catcode `\%12\relax}%
\providecommand \@@startlink[1]{}%
\providecommand \@@endlink[0]{}%
\providecommand \url  [0]{\begingroup\@sanitize@url \@url }%
\providecommand \@url [1]{\endgroup\@href {#1}{\urlprefix }}%
\providecommand \urlprefix  [0]{URL }%
\providecommand \Eprint [0]{\href }%
\providecommand \doibase [0]{https://doi.org/}%
\providecommand \selectlanguage [0]{\@gobble}%
\providecommand \bibinfo  [0]{\@secondoftwo}%
\providecommand \bibfield  [0]{\@secondoftwo}%
\providecommand \translation [1]{[#1]}%
\providecommand \BibitemOpen [0]{}%
\providecommand \bibitemStop [0]{}%
\providecommand \bibitemNoStop [0]{.\EOS\space}%
\providecommand \EOS [0]{\spacefactor3000\relax}%
\providecommand \BibitemShut  [1]{\csname bibitem#1\endcsname}%
\let\auto@bib@innerbib\@empty
\bibitem [{\citenamefont {Mitra}\ \emph {et~al.}(2020)\citenamefont {Mitra},
  \citenamefont {Martin}, \citenamefont {Biedermann}, \citenamefont {Marino},
  \citenamefont {Poggi},\ and\ \citenamefont {Deutsch}}]{mitra2020robust}%
  \BibitemOpen
  \bibfield  {author} {\bibinfo {author} {\bibfnamefont {A.}~\bibnamefont
  {Mitra}}, \bibinfo {author} {\bibfnamefont {M.~J.}\ \bibnamefont {Martin}},
  \bibinfo {author} {\bibfnamefont {G.~W.}\ \bibnamefont {Biedermann}},
  \bibinfo {author} {\bibfnamefont {A.~M.}\ \bibnamefont {Marino}}, \bibinfo
  {author} {\bibfnamefont {P.~M.}\ \bibnamefont {Poggi}},\ and\ \bibinfo
  {author} {\bibfnamefont {I.~H.}\ \bibnamefont {Deutsch}},\ }\bibfield
  {title} {\bibinfo {title} {Robust m{\o}lmer-s{\o}rensen gate for neutral
  atoms using rapid adiabatic rydberg dressing},\ }\href@noop {} {\bibfield
  {journal} {\bibinfo  {journal} {Phys. Rev. A}\ }\textbf {\bibinfo {volume}
  {101}},\ \bibinfo {pages} {030301 (R)} (\bibinfo {year} {2020})}\BibitemShut
  {NoStop}%
\bibitem [{\citenamefont {Brennen}\ \emph {et~al.}(1999)\citenamefont
  {Brennen}, \citenamefont {Caves}, \citenamefont {Jessen},\ and\ \citenamefont
  {Deutsch}}]{brennen1999quantum}%
  \BibitemOpen
  \bibfield  {author} {\bibinfo {author} {\bibfnamefont {G.~K.}\ \bibnamefont
  {Brennen}}, \bibinfo {author} {\bibfnamefont {C.~M.}\ \bibnamefont {Caves}},
  \bibinfo {author} {\bibfnamefont {P.~S.}\ \bibnamefont {Jessen}},\ and\
  \bibinfo {author} {\bibfnamefont {I.~H.}\ \bibnamefont {Deutsch}},\
  }\bibfield  {title} {\bibinfo {title} {Quantum logic gates in optical
  lattices},\ }\href@noop {} {\bibfield  {journal} {\bibinfo  {journal} {Phys.
  Rev. Lett.}\ }\textbf {\bibinfo {volume} {82}},\ \bibinfo {pages} {1060}
  (\bibinfo {year} {1999})}\BibitemShut {NoStop}%
\bibitem [{\citenamefont {Isenhower}\ \emph {et~al.}(2010)\citenamefont
  {Isenhower}, \citenamefont {Urban}, \citenamefont {Zhang}, \citenamefont
  {Gill}, \citenamefont {Henage}, \citenamefont {Johnson}, \citenamefont
  {Walker},\ and\ \citenamefont {Saffman}}]{isenhower2010demonstration}%
  \BibitemOpen
  \bibfield  {author} {\bibinfo {author} {\bibfnamefont {L.}~\bibnamefont
  {Isenhower}}, \bibinfo {author} {\bibfnamefont {E.}~\bibnamefont {Urban}},
  \bibinfo {author} {\bibfnamefont {X.-L.}\ \bibnamefont {Zhang}}, \bibinfo
  {author} {\bibfnamefont {A.}~\bibnamefont {Gill}}, \bibinfo {author}
  {\bibfnamefont {T.}~\bibnamefont {Henage}}, \bibinfo {author} {\bibfnamefont
  {T.-A.}\ \bibnamefont {Johnson}}, \bibinfo {author} {\bibfnamefont {T.-G.}\
  \bibnamefont {Walker}},\ and\ \bibinfo {author} {\bibfnamefont
  {M.}~\bibnamefont {Saffman}},\ }\bibfield  {title} {\bibinfo {title}
  {Demonstration of a neutral atom controlled-not quantum gate},\ }\href@noop
  {} {\bibfield  {journal} {\bibinfo  {journal} {Phys. Rev. Lett.}\ }\textbf
  {\bibinfo {volume} {104}},\ \bibinfo {pages} {010503} (\bibinfo {year}
  {2010})}\BibitemShut {NoStop}%
\bibitem [{\citenamefont {Saffman}\ \emph {et~al.}(2010)\citenamefont
  {Saffman}, \citenamefont {Walker},\ and\ \citenamefont
  {M{\o}lmer}}]{saffman2010quantum}%
  \BibitemOpen
  \bibfield  {author} {\bibinfo {author} {\bibfnamefont {M.}~\bibnamefont
  {Saffman}}, \bibinfo {author} {\bibfnamefont {T.~G.}\ \bibnamefont
  {Walker}},\ and\ \bibinfo {author} {\bibfnamefont {K.}~\bibnamefont
  {M{\o}lmer}},\ }\bibfield  {title} {\bibinfo {title} {Quantum information
  with rydberg atoms},\ }\href@noop {} {\bibfield  {journal} {\bibinfo
  {journal} {Reviews of Modern Physics}\ }\textbf {\bibinfo {volume} {82}},\
  \bibinfo {pages} {2313} (\bibinfo {year} {2010})}\BibitemShut {NoStop}%
\bibitem [{\citenamefont {Saffman}(2016)}]{saffman2016quantum}%
  \BibitemOpen
  \bibfield  {author} {\bibinfo {author} {\bibfnamefont {M.}~\bibnamefont
  {Saffman}},\ }\bibfield  {title} {\bibinfo {title} {Quantum computing with
  atomic qubits and rydberg interactions: progress and challenges},\
  }\href@noop {} {\bibfield  {journal} {\bibinfo  {journal} {Journal of Physics
  B: Atomic, Molecular and Optical Physics}\ }\textbf {\bibinfo {volume}
  {49}},\ \bibinfo {pages} {202001} (\bibinfo {year} {2016})}\BibitemShut
  {NoStop}%
\bibitem [{\citenamefont {Levine}\ \emph {et~al.}(2019)\citenamefont {Levine},
  \citenamefont {Keesling}, \citenamefont {Semeghini}, \citenamefont {Omran},
  \citenamefont {Wang}, \citenamefont {Ebadi}, \citenamefont {Bernien},
  \citenamefont {Greiner}, \citenamefont {Vuleti{\'c}}, \citenamefont {Pichler}
  \emph {et~al.}}]{levine2019parallel}%
  \BibitemOpen
  \bibfield  {author} {\bibinfo {author} {\bibfnamefont {H.}~\bibnamefont
  {Levine}}, \bibinfo {author} {\bibfnamefont {A.}~\bibnamefont {Keesling}},
  \bibinfo {author} {\bibfnamefont {G.}~\bibnamefont {Semeghini}}, \bibinfo
  {author} {\bibfnamefont {A.}~\bibnamefont {Omran}}, \bibinfo {author}
  {\bibfnamefont {T.~T.}\ \bibnamefont {Wang}}, \bibinfo {author}
  {\bibfnamefont {S.}~\bibnamefont {Ebadi}}, \bibinfo {author} {\bibfnamefont
  {H.}~\bibnamefont {Bernien}}, \bibinfo {author} {\bibfnamefont
  {M.}~\bibnamefont {Greiner}}, \bibinfo {author} {\bibfnamefont
  {V.}~\bibnamefont {Vuleti{\'c}}}, \bibinfo {author} {\bibfnamefont
  {H.}~\bibnamefont {Pichler}}, \emph {et~al.},\ }\bibfield  {title} {\bibinfo
  {title} {Parallel implementation of high-fidelity multi-qubit gates with
  neutral atoms},\ }\href@noop {} {\bibfield  {journal} {\bibinfo  {journal}
  {Phys. Rev. Let}\ }\textbf {\bibinfo {volume} {123}},\ \bibinfo {pages}
  {170503} (\bibinfo {year} {2019})}\BibitemShut {NoStop}%
\bibitem [{\citenamefont {Bluvstein}\ \emph {et~al.}(2022)\citenamefont
  {Bluvstein}, \citenamefont {Levine}, \citenamefont {Semeghini}, \citenamefont
  {Wang}, \citenamefont {Ebadi}, \citenamefont {Kalinowski}, \citenamefont
  {Keesling}, \citenamefont {Maskara}, \citenamefont {Pichler}, \citenamefont
  {Greiner} \emph {et~al.}}]{bluvstein2022quantum}%
  \BibitemOpen
  \bibfield  {author} {\bibinfo {author} {\bibfnamefont {D.}~\bibnamefont
  {Bluvstein}}, \bibinfo {author} {\bibfnamefont {H.}~\bibnamefont {Levine}},
  \bibinfo {author} {\bibfnamefont {G.}~\bibnamefont {Semeghini}}, \bibinfo
  {author} {\bibfnamefont {T.~T.}\ \bibnamefont {Wang}}, \bibinfo {author}
  {\bibfnamefont {S.}~\bibnamefont {Ebadi}}, \bibinfo {author} {\bibfnamefont
  {M.}~\bibnamefont {Kalinowski}}, \bibinfo {author} {\bibfnamefont
  {A.}~\bibnamefont {Keesling}}, \bibinfo {author} {\bibfnamefont
  {N.}~\bibnamefont {Maskara}}, \bibinfo {author} {\bibfnamefont
  {H.}~\bibnamefont {Pichler}}, \bibinfo {author} {\bibfnamefont
  {M.}~\bibnamefont {Greiner}}, \emph {et~al.},\ }\bibfield  {title} {\bibinfo
  {title} {A quantum processor based on coherent transport of entangled atom
  arrays},\ }\href@noop {} {\bibfield  {journal} {\bibinfo  {journal} {Nature}\
  }\textbf {\bibinfo {volume} {604}},\ \bibinfo {pages} {451} (\bibinfo {year}
  {2022})}\BibitemShut {NoStop}%
\bibitem [{\citenamefont {Graham}\ \emph {et~al.}(2022)\citenamefont {Graham},
  \citenamefont {Song}, \citenamefont {Scott}, \citenamefont {Poole},
  \citenamefont {Phuttitarn}, \citenamefont {Jooya}, \citenamefont {Eichler},
  \citenamefont {Jiang}, \citenamefont {Marra}, \citenamefont {Grinkemeyer}
  \emph {et~al.}}]{graham2022multi}%
  \BibitemOpen
  \bibfield  {author} {\bibinfo {author} {\bibfnamefont {T.}~\bibnamefont
  {Graham}}, \bibinfo {author} {\bibfnamefont {Y.}~\bibnamefont {Song}},
  \bibinfo {author} {\bibfnamefont {J.}~\bibnamefont {Scott}}, \bibinfo
  {author} {\bibfnamefont {C.}~\bibnamefont {Poole}}, \bibinfo {author}
  {\bibfnamefont {L.}~\bibnamefont {Phuttitarn}}, \bibinfo {author}
  {\bibfnamefont {K.}~\bibnamefont {Jooya}}, \bibinfo {author} {\bibfnamefont
  {P.}~\bibnamefont {Eichler}}, \bibinfo {author} {\bibfnamefont
  {X.}~\bibnamefont {Jiang}}, \bibinfo {author} {\bibfnamefont
  {A.}~\bibnamefont {Marra}}, \bibinfo {author} {\bibfnamefont
  {B.}~\bibnamefont {Grinkemeyer}}, \emph {et~al.},\ }\bibfield  {title}
  {\bibinfo {title} {Multi-qubit entanglement and algorithms on a neutral-atom
  quantum computer},\ }\href@noop {} {\bibfield  {journal} {\bibinfo  {journal}
  {Nature}\ }\textbf {\bibinfo {volume} {604}},\ \bibinfo {pages} {457}
  (\bibinfo {year} {2022})}\BibitemShut {NoStop}%
\bibitem [{\citenamefont {Ebadi}\ \emph {et~al.}(2022)\citenamefont {Ebadi},
  \citenamefont {Keesling}, \citenamefont {Cain}, \citenamefont {Wang},
  \citenamefont {Levine}, \citenamefont {Bluvstein}, \citenamefont {Semeghini},
  \citenamefont {Omran}, \citenamefont {Liu}, \citenamefont {Samajdar} \emph
  {et~al.}}]{ebadi2022quantum}%
  \BibitemOpen
  \bibfield  {author} {\bibinfo {author} {\bibfnamefont {S.}~\bibnamefont
  {Ebadi}}, \bibinfo {author} {\bibfnamefont {A.}~\bibnamefont {Keesling}},
  \bibinfo {author} {\bibfnamefont {M.}~\bibnamefont {Cain}}, \bibinfo {author}
  {\bibfnamefont {T.~T.}\ \bibnamefont {Wang}}, \bibinfo {author}
  {\bibfnamefont {H.}~\bibnamefont {Levine}}, \bibinfo {author} {\bibfnamefont
  {D.}~\bibnamefont {Bluvstein}}, \bibinfo {author} {\bibfnamefont
  {G.}~\bibnamefont {Semeghini}}, \bibinfo {author} {\bibfnamefont
  {A.}~\bibnamefont {Omran}}, \bibinfo {author} {\bibfnamefont {J.-G.}\
  \bibnamefont {Liu}}, \bibinfo {author} {\bibfnamefont {R.}~\bibnamefont
  {Samajdar}}, \emph {et~al.},\ }\bibfield  {title} {\bibinfo {title} {Quantum
  optimization of maximum independent set using rydberg atom arrays},\
  }\href@noop {} {\bibfield  {journal} {\bibinfo  {journal} {Science}\ ,\
  \bibinfo {pages} {eabo6587}} (\bibinfo {year} {2022})}\BibitemShut {NoStop}%
\bibitem [{\citenamefont {Zeiher}\ \emph {et~al.}(2016)\citenamefont {Zeiher},
  \citenamefont {Van~Bijnen}, \citenamefont {Schau{\ss}}, \citenamefont {Hild},
  \citenamefont {Choi}, \citenamefont {Pohl}, \citenamefont {Bloch},\ and\
  \citenamefont {Gross}}]{zeiher2016many}%
  \BibitemOpen
  \bibfield  {author} {\bibinfo {author} {\bibfnamefont {J.}~\bibnamefont
  {Zeiher}}, \bibinfo {author} {\bibfnamefont {R.}~\bibnamefont {Van~Bijnen}},
  \bibinfo {author} {\bibfnamefont {P.}~\bibnamefont {Schau{\ss}}}, \bibinfo
  {author} {\bibfnamefont {S.}~\bibnamefont {Hild}}, \bibinfo {author}
  {\bibfnamefont {J.-y.}\ \bibnamefont {Choi}}, \bibinfo {author}
  {\bibfnamefont {T.}~\bibnamefont {Pohl}}, \bibinfo {author} {\bibfnamefont
  {I.}~\bibnamefont {Bloch}},\ and\ \bibinfo {author} {\bibfnamefont
  {C.}~\bibnamefont {Gross}},\ }\bibfield  {title} {\bibinfo {title} {Many-body
  interferometry of a rydberg-dressed spin lattice},\ }\href@noop {} {\bibfield
   {journal} {\bibinfo  {journal} {Nature Physics}\ }\textbf {\bibinfo {volume}
  {12}},\ \bibinfo {pages} {1095} (\bibinfo {year} {2016})}\BibitemShut
  {NoStop}%
\bibitem [{\citenamefont {Zeiher}\ \emph {et~al.}(2017)\citenamefont {Zeiher},
  \citenamefont {Choi}, \citenamefont {Rubio-Abadal}, \citenamefont {Pohl},
  \citenamefont {Van~Bijnen}, \citenamefont {Bloch},\ and\ \citenamefont
  {Gross}}]{zeiher2017coherent}%
  \BibitemOpen
  \bibfield  {author} {\bibinfo {author} {\bibfnamefont {J.}~\bibnamefont
  {Zeiher}}, \bibinfo {author} {\bibfnamefont {J.-y.}\ \bibnamefont {Choi}},
  \bibinfo {author} {\bibfnamefont {A.}~\bibnamefont {Rubio-Abadal}}, \bibinfo
  {author} {\bibfnamefont {T.}~\bibnamefont {Pohl}}, \bibinfo {author}
  {\bibfnamefont {R.}~\bibnamefont {Van~Bijnen}}, \bibinfo {author}
  {\bibfnamefont {I.}~\bibnamefont {Bloch}},\ and\ \bibinfo {author}
  {\bibfnamefont {C.}~\bibnamefont {Gross}},\ }\bibfield  {title} {\bibinfo
  {title} {Coherent many-body spin dynamics in a long-range interacting ising
  chain},\ }\href@noop {} {\bibfield  {journal} {\bibinfo  {journal} {Phys.
  Rev. X}\ }\textbf {\bibinfo {volume} {7}},\ \bibinfo {pages} {041063}
  (\bibinfo {year} {2017})}\BibitemShut {NoStop}%
\bibitem [{\citenamefont {Borish}\ \emph {et~al.}(2020)\citenamefont {Borish},
  \citenamefont {Markovi{\'c}}, \citenamefont {Hines}, \citenamefont
  {Rajagopal},\ and\ \citenamefont {Schleier-Smith}}]{borish2020transverse}%
  \BibitemOpen
  \bibfield  {author} {\bibinfo {author} {\bibfnamefont {V.}~\bibnamefont
  {Borish}}, \bibinfo {author} {\bibfnamefont {O.}~\bibnamefont
  {Markovi{\'c}}}, \bibinfo {author} {\bibfnamefont {J.~A.}\ \bibnamefont
  {Hines}}, \bibinfo {author} {\bibfnamefont {S.~V.}\ \bibnamefont
  {Rajagopal}},\ and\ \bibinfo {author} {\bibfnamefont {M.}~\bibnamefont
  {Schleier-Smith}},\ }\bibfield  {title} {\bibinfo {title} {Transverse-field
  ising dynamics in a rydberg-dressed atomic gas},\ }\href@noop {} {\bibfield
  {journal} {\bibinfo  {journal} {Phys. Rev. Lett.}\ }\textbf {\bibinfo
  {volume} {124}},\ \bibinfo {pages} {063601} (\bibinfo {year}
  {2020})}\BibitemShut {NoStop}%
\bibitem [{\citenamefont {Browaeys}\ and\ \citenamefont
  {Lahaye}(2020)}]{browaeys2020many}%
  \BibitemOpen
  \bibfield  {author} {\bibinfo {author} {\bibfnamefont {A.}~\bibnamefont
  {Browaeys}}\ and\ \bibinfo {author} {\bibfnamefont {T.}~\bibnamefont
  {Lahaye}},\ }\bibfield  {title} {\bibinfo {title} {Many-body physics with
  individually controlled rydberg atoms},\ }\href@noop {} {\bibfield  {journal}
  {\bibinfo  {journal} {Nature Physics}\ }\textbf {\bibinfo {volume} {16}},\
  \bibinfo {pages} {132} (\bibinfo {year} {2020})}\BibitemShut {NoStop}%
\bibitem [{\citenamefont {Bernien}\ \emph {et~al.}(2017)\citenamefont
  {Bernien}, \citenamefont {Schwartz}, \citenamefont {Keesling}, \citenamefont
  {Levine}, \citenamefont {Omran}, \citenamefont {Pichler}, \citenamefont
  {Choi}, \citenamefont {Zibrov}, \citenamefont {Endres}, \citenamefont
  {Greiner} \emph {et~al.}}]{bernien2017probing}%
  \BibitemOpen
  \bibfield  {author} {\bibinfo {author} {\bibfnamefont {H.}~\bibnamefont
  {Bernien}}, \bibinfo {author} {\bibfnamefont {S.}~\bibnamefont {Schwartz}},
  \bibinfo {author} {\bibfnamefont {A.}~\bibnamefont {Keesling}}, \bibinfo
  {author} {\bibfnamefont {H.}~\bibnamefont {Levine}}, \bibinfo {author}
  {\bibfnamefont {A.}~\bibnamefont {Omran}}, \bibinfo {author} {\bibfnamefont
  {H.}~\bibnamefont {Pichler}}, \bibinfo {author} {\bibfnamefont
  {S.}~\bibnamefont {Choi}}, \bibinfo {author} {\bibfnamefont {A.~S.}\
  \bibnamefont {Zibrov}}, \bibinfo {author} {\bibfnamefont {M.}~\bibnamefont
  {Endres}}, \bibinfo {author} {\bibfnamefont {M.}~\bibnamefont {Greiner}},
  \emph {et~al.},\ }\bibfield  {title} {\bibinfo {title} {Probing many-body
  dynamics on a 51-atom quantum simulator},\ }\href@noop {} {\bibfield
  {journal} {\bibinfo  {journal} {Nature}\ }\textbf {\bibinfo {volume} {551}},\
  \bibinfo {pages} {579} (\bibinfo {year} {2017})}\BibitemShut {NoStop}%
\bibitem [{\citenamefont {Keesling}\ \emph {et~al.}(2019)\citenamefont
  {Keesling}, \citenamefont {Omran}, \citenamefont {Levine}, \citenamefont
  {Bernien}, \citenamefont {Pichler}, \citenamefont {Choi}, \citenamefont
  {Samajdar}, \citenamefont {Schwartz}, \citenamefont {Silvi}, \citenamefont
  {Sachdev} \emph {et~al.}}]{keesling2019quantum}%
  \BibitemOpen
  \bibfield  {author} {\bibinfo {author} {\bibfnamefont {A.}~\bibnamefont
  {Keesling}}, \bibinfo {author} {\bibfnamefont {A.}~\bibnamefont {Omran}},
  \bibinfo {author} {\bibfnamefont {H.}~\bibnamefont {Levine}}, \bibinfo
  {author} {\bibfnamefont {H.}~\bibnamefont {Bernien}}, \bibinfo {author}
  {\bibfnamefont {H.}~\bibnamefont {Pichler}}, \bibinfo {author} {\bibfnamefont
  {S.}~\bibnamefont {Choi}}, \bibinfo {author} {\bibfnamefont {R.}~\bibnamefont
  {Samajdar}}, \bibinfo {author} {\bibfnamefont {S.}~\bibnamefont {Schwartz}},
  \bibinfo {author} {\bibfnamefont {P.}~\bibnamefont {Silvi}}, \bibinfo
  {author} {\bibfnamefont {S.}~\bibnamefont {Sachdev}}, \emph {et~al.},\
  }\bibfield  {title} {\bibinfo {title} {Quantum kibble--zurek mechanism and
  critical dynamics on a programmable rydberg simulator},\ }\href@noop {}
  {\bibfield  {journal} {\bibinfo  {journal} {Nature}\ }\textbf {\bibinfo
  {volume} {568}},\ \bibinfo {pages} {207} (\bibinfo {year}
  {2019})}\BibitemShut {NoStop}%
\bibitem [{\citenamefont {Ebadi}\ \emph {et~al.}(2021)\citenamefont {Ebadi},
  \citenamefont {Wang}, \citenamefont {Levine}, \citenamefont {Keesling},
  \citenamefont {Semeghini}, \citenamefont {Omran}, \citenamefont {Bluvstein},
  \citenamefont {Samajdar}, \citenamefont {Pichler}, \citenamefont {Ho} \emph
  {et~al.}}]{ebadi2021quantum}%
  \BibitemOpen
  \bibfield  {author} {\bibinfo {author} {\bibfnamefont {S.}~\bibnamefont
  {Ebadi}}, \bibinfo {author} {\bibfnamefont {T.~T.}\ \bibnamefont {Wang}},
  \bibinfo {author} {\bibfnamefont {H.}~\bibnamefont {Levine}}, \bibinfo
  {author} {\bibfnamefont {A.}~\bibnamefont {Keesling}}, \bibinfo {author}
  {\bibfnamefont {G.}~\bibnamefont {Semeghini}}, \bibinfo {author}
  {\bibfnamefont {A.}~\bibnamefont {Omran}}, \bibinfo {author} {\bibfnamefont
  {D.}~\bibnamefont {Bluvstein}}, \bibinfo {author} {\bibfnamefont
  {R.}~\bibnamefont {Samajdar}}, \bibinfo {author} {\bibfnamefont
  {H.}~\bibnamefont {Pichler}}, \bibinfo {author} {\bibfnamefont {W.~W.}\
  \bibnamefont {Ho}}, \emph {et~al.},\ }\bibfield  {title} {\bibinfo {title}
  {Quantum phases of matter on a 256-atom programmable quantum simulator},\
  }\href@noop {} {\bibfield  {journal} {\bibinfo  {journal} {Nature}\ }\textbf
  {\bibinfo {volume} {595}},\ \bibinfo {pages} {227} (\bibinfo {year}
  {2021})}\BibitemShut {NoStop}%
\bibitem [{\citenamefont {Scholl}\ \emph {et~al.}(2021)\citenamefont {Scholl},
  \citenamefont {Schuler}, \citenamefont {Williams}, \citenamefont
  {Eberharter}, \citenamefont {Barredo}, \citenamefont {Schymik}, \citenamefont
  {Lienhard}, \citenamefont {Henry}, \citenamefont {Lang}, \citenamefont
  {Lahaye} \emph {et~al.}}]{scholl2021quantum}%
  \BibitemOpen
  \bibfield  {author} {\bibinfo {author} {\bibfnamefont {P.}~\bibnamefont
  {Scholl}}, \bibinfo {author} {\bibfnamefont {M.}~\bibnamefont {Schuler}},
  \bibinfo {author} {\bibfnamefont {H.~J.}\ \bibnamefont {Williams}}, \bibinfo
  {author} {\bibfnamefont {A.~A.}\ \bibnamefont {Eberharter}}, \bibinfo
  {author} {\bibfnamefont {D.}~\bibnamefont {Barredo}}, \bibinfo {author}
  {\bibfnamefont {K.-N.}\ \bibnamefont {Schymik}}, \bibinfo {author}
  {\bibfnamefont {V.}~\bibnamefont {Lienhard}}, \bibinfo {author}
  {\bibfnamefont {L.-P.}\ \bibnamefont {Henry}}, \bibinfo {author}
  {\bibfnamefont {T.~C.}\ \bibnamefont {Lang}}, \bibinfo {author}
  {\bibfnamefont {T.}~\bibnamefont {Lahaye}}, \emph {et~al.},\ }\bibfield
  {title} {\bibinfo {title} {Quantum simulation of 2d antiferromagnets with
  hundreds of rydberg atoms},\ }\href@noop {} {\bibfield  {journal} {\bibinfo
  {journal} {Nature}\ }\textbf {\bibinfo {volume} {595}},\ \bibinfo {pages}
  {233} (\bibinfo {year} {2021})}\BibitemShut {NoStop}%
\bibitem [{\citenamefont {Kaubruegger}\ \emph {et~al.}(2019)\citenamefont
  {Kaubruegger}, \citenamefont {Silvi}, \citenamefont {Kokail}, \citenamefont
  {van Bijnen}, \citenamefont {Rey}, \citenamefont {Ye}, \citenamefont
  {Kaufman},\ and\ \citenamefont {Zoller}}]{kaubruegger2019variational}%
  \BibitemOpen
  \bibfield  {author} {\bibinfo {author} {\bibfnamefont {R.}~\bibnamefont
  {Kaubruegger}}, \bibinfo {author} {\bibfnamefont {P.}~\bibnamefont {Silvi}},
  \bibinfo {author} {\bibfnamefont {C.}~\bibnamefont {Kokail}}, \bibinfo
  {author} {\bibfnamefont {R.}~\bibnamefont {van Bijnen}}, \bibinfo {author}
  {\bibfnamefont {A.~M.}\ \bibnamefont {Rey}}, \bibinfo {author} {\bibfnamefont
  {J.}~\bibnamefont {Ye}}, \bibinfo {author} {\bibfnamefont {A.~M.}\
  \bibnamefont {Kaufman}},\ and\ \bibinfo {author} {\bibfnamefont
  {P.}~\bibnamefont {Zoller}},\ }\bibfield  {title} {\bibinfo {title}
  {Variational spin-squeezing algorithms on programmable quantum sensors},\
  }\href@noop {} {\bibfield  {journal} {\bibinfo  {journal} {Phys. Rev. Lett.}\
  }\textbf {\bibinfo {volume} {123}},\ \bibinfo {pages} {260505} (\bibinfo
  {year} {2019})}\BibitemShut {NoStop}%
\bibitem [{\citenamefont {Van~Damme}\ \emph {et~al.}(2021)\citenamefont
  {Van~Damme}, \citenamefont {Zheng}, \citenamefont {Saffman}, \citenamefont
  {Vavilov},\ and\ \citenamefont {Kolkowitz}}]{van2021impacts}%
  \BibitemOpen
  \bibfield  {author} {\bibinfo {author} {\bibfnamefont {J.}~\bibnamefont
  {Van~Damme}}, \bibinfo {author} {\bibfnamefont {X.}~\bibnamefont {Zheng}},
  \bibinfo {author} {\bibfnamefont {M.}~\bibnamefont {Saffman}}, \bibinfo
  {author} {\bibfnamefont {M.~G.}\ \bibnamefont {Vavilov}},\ and\ \bibinfo
  {author} {\bibfnamefont {S.}~\bibnamefont {Kolkowitz}},\ }\bibfield  {title}
  {\bibinfo {title} {Impacts of random filling on spin squeezing via rydberg
  dressing in optical clocks},\ }\href@noop {} {\bibfield  {journal} {\bibinfo
  {journal} {Phys. Rev. A}\ }\textbf {\bibinfo {volume} {103}},\ \bibinfo
  {pages} {023106} (\bibinfo {year} {2021})}\BibitemShut {NoStop}%
\bibitem [{\citenamefont {Schine}\ \emph {et~al.}(2022)\citenamefont {Schine},
  \citenamefont {Young}, \citenamefont {Eckner}, \citenamefont {Martin},\ and\
  \citenamefont {Kaufman}}]{schine2022long}%
  \BibitemOpen
  \bibfield  {author} {\bibinfo {author} {\bibfnamefont {N.}~\bibnamefont
  {Schine}}, \bibinfo {author} {\bibfnamefont {A.~W.}\ \bibnamefont {Young}},
  \bibinfo {author} {\bibfnamefont {W.~J.}\ \bibnamefont {Eckner}}, \bibinfo
  {author} {\bibfnamefont {M.~J.}\ \bibnamefont {Martin}},\ and\ \bibinfo
  {author} {\bibfnamefont {A.~M.}\ \bibnamefont {Kaufman}},\ }\bibfield
  {title} {\bibinfo {title} {Long-lived bell states in an array of optical
  clock qubits},\ }\href@noop {} {\bibfield  {journal} {\bibinfo  {journal}
  {Nature Physics}\ }\textbf {\bibinfo {volume} {18}},\ \bibinfo {pages} {1067}
  (\bibinfo {year} {2022})}\BibitemShut {NoStop}%
\bibitem [{\citenamefont {Jaksch}\ \emph {et~al.}(2000)\citenamefont {Jaksch},
  \citenamefont {Cirac}, \citenamefont {Zoller}, \citenamefont {Rolston},
  \citenamefont {C{\^o}t{\'e}},\ and\ \citenamefont {Lukin}}]{jaksch2000fast}%
  \BibitemOpen
  \bibfield  {author} {\bibinfo {author} {\bibfnamefont {D.}~\bibnamefont
  {Jaksch}}, \bibinfo {author} {\bibfnamefont {J.}~\bibnamefont {Cirac}},
  \bibinfo {author} {\bibfnamefont {P.}~\bibnamefont {Zoller}}, \bibinfo
  {author} {\bibfnamefont {S.}~\bibnamefont {Rolston}}, \bibinfo {author}
  {\bibfnamefont {R.}~\bibnamefont {C{\^o}t{\'e}}},\ and\ \bibinfo {author}
  {\bibfnamefont {M.}~\bibnamefont {Lukin}},\ }\bibfield  {title} {\bibinfo
  {title} {Fast quantum gates for neutral atoms},\ }\href@noop {} {\bibfield
  {journal} {\bibinfo  {journal} {Phys. Rev. Lett.}\ }\textbf {\bibinfo
  {volume} {85}},\ \bibinfo {pages} {2208} (\bibinfo {year}
  {2000})}\BibitemShut {NoStop}%
\bibitem [{\citenamefont {Beterov}\ \emph {et~al.}(2013)\citenamefont
  {Beterov}, \citenamefont {Saffman}, \citenamefont {Yakshina}, \citenamefont
  {Zhukov}, \citenamefont {Tretyakov}, \citenamefont {Entin}, \citenamefont
  {Ryabtsev}, \citenamefont {Mansell}, \citenamefont {MacCormick},
  \citenamefont {Bergamini} \emph {et~al.}}]{beterov2013quantum}%
  \BibitemOpen
  \bibfield  {author} {\bibinfo {author} {\bibfnamefont {I.}~\bibnamefont
  {Beterov}}, \bibinfo {author} {\bibfnamefont {M.}~\bibnamefont {Saffman}},
  \bibinfo {author} {\bibfnamefont {E.}~\bibnamefont {Yakshina}}, \bibinfo
  {author} {\bibfnamefont {V.}~\bibnamefont {Zhukov}}, \bibinfo {author}
  {\bibfnamefont {D.}~\bibnamefont {Tretyakov}}, \bibinfo {author}
  {\bibfnamefont {V.}~\bibnamefont {Entin}}, \bibinfo {author} {\bibfnamefont
  {I.}~\bibnamefont {Ryabtsev}}, \bibinfo {author} {\bibfnamefont
  {C.}~\bibnamefont {Mansell}}, \bibinfo {author} {\bibfnamefont
  {C.}~\bibnamefont {MacCormick}}, \bibinfo {author} {\bibfnamefont
  {S.}~\bibnamefont {Bergamini}}, \emph {et~al.},\ }\bibfield  {title}
  {\bibinfo {title} {Quantum gates in mesoscopic atomic ensembles based on
  adiabatic passage and rydberg blockade},\ }\href@noop {} {\bibfield
  {journal} {\bibinfo  {journal} {Phys. Rev. A}\ }\textbf {\bibinfo {volume}
  {88}},\ \bibinfo {pages} {010303} (\bibinfo {year} {2013})}\BibitemShut
  {NoStop}%
\bibitem [{\citenamefont {Keating}\ \emph {et~al.}(2015)\citenamefont
  {Keating}, \citenamefont {Cook}, \citenamefont {Hankin}, \citenamefont {Jau},
  \citenamefont {Biedermann},\ and\ \citenamefont
  {Deutsch}}]{keating2015robust}%
  \BibitemOpen
  \bibfield  {author} {\bibinfo {author} {\bibfnamefont {T.}~\bibnamefont
  {Keating}}, \bibinfo {author} {\bibfnamefont {R.~L.}\ \bibnamefont {Cook}},
  \bibinfo {author} {\bibfnamefont {A.~M.}\ \bibnamefont {Hankin}}, \bibinfo
  {author} {\bibfnamefont {Y.-Y.}\ \bibnamefont {Jau}}, \bibinfo {author}
  {\bibfnamefont {G.~W.}\ \bibnamefont {Biedermann}},\ and\ \bibinfo {author}
  {\bibfnamefont {I.~H.}\ \bibnamefont {Deutsch}},\ }\bibfield  {title}
  {\bibinfo {title} {Robust quantum logic in neutral atoms via adiabatic
  rydberg dressing},\ }\href@noop {} {\bibfield  {journal} {\bibinfo  {journal}
  {Phys. Rev. A}\ }\textbf {\bibinfo {volume} {91}},\ \bibinfo {pages} {012337}
  (\bibinfo {year} {2015})}\BibitemShut {NoStop}%
\bibitem [{\citenamefont {Beterov}\ \emph {et~al.}(2016)\citenamefont
  {Beterov}, \citenamefont {Saffman}, \citenamefont {Yakshina}, \citenamefont
  {Tretyakov}, \citenamefont {Entin}, \citenamefont {Bergamini}, \citenamefont
  {Kuznetsova},\ and\ \citenamefont {Ryabtsev}}]{beterov2016two}%
  \BibitemOpen
  \bibfield  {author} {\bibinfo {author} {\bibfnamefont {I.}~\bibnamefont
  {Beterov}}, \bibinfo {author} {\bibfnamefont {M.}~\bibnamefont {Saffman}},
  \bibinfo {author} {\bibfnamefont {E.}~\bibnamefont {Yakshina}}, \bibinfo
  {author} {\bibfnamefont {D.}~\bibnamefont {Tretyakov}}, \bibinfo {author}
  {\bibfnamefont {V.}~\bibnamefont {Entin}}, \bibinfo {author} {\bibfnamefont
  {S.}~\bibnamefont {Bergamini}}, \bibinfo {author} {\bibfnamefont
  {E.}~\bibnamefont {Kuznetsova}},\ and\ \bibinfo {author} {\bibfnamefont
  {I.}~\bibnamefont {Ryabtsev}},\ }\bibfield  {title} {\bibinfo {title}
  {Two-qubit gates using adiabatic passage of the stark-tuned f{\"o}rster
  resonances in rydberg atoms},\ }\href@noop {} {\bibfield  {journal} {\bibinfo
   {journal} {Phys. Rev. A}\ }\textbf {\bibinfo {volume} {94}},\ \bibinfo
  {pages} {062307} (\bibinfo {year} {2016})}\BibitemShut {NoStop}%
\bibitem [{\citenamefont {Beterov}\ \emph {et~al.}(2018)\citenamefont
  {Beterov}, \citenamefont {Hamzina}, \citenamefont {Yakshina}, \citenamefont
  {Tretyakov}, \citenamefont {Entin},\ and\ \citenamefont
  {Ryabtsev}}]{beterov2018adiabatic}%
  \BibitemOpen
  \bibfield  {author} {\bibinfo {author} {\bibfnamefont {I.}~\bibnamefont
  {Beterov}}, \bibinfo {author} {\bibfnamefont {G.}~\bibnamefont {Hamzina}},
  \bibinfo {author} {\bibfnamefont {E.}~\bibnamefont {Yakshina}}, \bibinfo
  {author} {\bibfnamefont {D.}~\bibnamefont {Tretyakov}}, \bibinfo {author}
  {\bibfnamefont {V.}~\bibnamefont {Entin}},\ and\ \bibinfo {author}
  {\bibfnamefont {I.}~\bibnamefont {Ryabtsev}},\ }\bibfield  {title} {\bibinfo
  {title} {Adiabatic passage of radio-frequency-assisted f{\"o}rster resonances
  in rydberg atoms for two-qubit gates and the generation of bell states},\
  }\href@noop {} {\bibfield  {journal} {\bibinfo  {journal} {Phys. Rev. A}\
  }\textbf {\bibinfo {volume} {97}},\ \bibinfo {pages} {032701} (\bibinfo
  {year} {2018})}\BibitemShut {NoStop}%
\bibitem [{\citenamefont {Saffman}\ \emph {et~al.}(2020)\citenamefont
  {Saffman}, \citenamefont {Beterov}, \citenamefont {Dalal}, \citenamefont
  {P{\'a}ez},\ and\ \citenamefont {Sanders}}]{saffman2020symmetric}%
  \BibitemOpen
  \bibfield  {author} {\bibinfo {author} {\bibfnamefont {M.}~\bibnamefont
  {Saffman}}, \bibinfo {author} {\bibfnamefont {I.}~\bibnamefont {Beterov}},
  \bibinfo {author} {\bibfnamefont {A.}~\bibnamefont {Dalal}}, \bibinfo
  {author} {\bibfnamefont {E.}~\bibnamefont {P{\'a}ez}},\ and\ \bibinfo
  {author} {\bibfnamefont {B.}~\bibnamefont {Sanders}},\ }\bibfield  {title}
  {\bibinfo {title} {Symmetric rydberg controlled-z gates with adiabatic
  pulses},\ }\href@noop {} {\bibfield  {journal} {\bibinfo  {journal} {Phys.
  Rev. A}\ }\textbf {\bibinfo {volume} {101}},\ \bibinfo {pages} {062309}
  (\bibinfo {year} {2020})}\BibitemShut {NoStop}%
\bibitem [{\citenamefont {Beterov}\ \emph {et~al.}(2020)\citenamefont
  {Beterov}, \citenamefont {Tretyakov}, \citenamefont {Entin}, \citenamefont
  {Yakshina}, \citenamefont {Ryabtsev}, \citenamefont {Saffman},\ and\
  \citenamefont {Bergamini}}]{beterov2020application}%
  \BibitemOpen
  \bibfield  {author} {\bibinfo {author} {\bibfnamefont {I.}~\bibnamefont
  {Beterov}}, \bibinfo {author} {\bibfnamefont {D.}~\bibnamefont {Tretyakov}},
  \bibinfo {author} {\bibfnamefont {V.}~\bibnamefont {Entin}}, \bibinfo
  {author} {\bibfnamefont {E.}~\bibnamefont {Yakshina}}, \bibinfo {author}
  {\bibfnamefont {I.}~\bibnamefont {Ryabtsev}}, \bibinfo {author}
  {\bibfnamefont {M.}~\bibnamefont {Saffman}},\ and\ \bibinfo {author}
  {\bibfnamefont {S.}~\bibnamefont {Bergamini}},\ }\bibfield  {title} {\bibinfo
  {title} {Application of adiabatic passage in rydberg atomic ensembles for
  quantum information processing},\ }\href@noop {} {\bibfield  {journal}
  {\bibinfo  {journal} {Journal of Physics B: Atomic, Molecular and Optical
  Physics}\ }\textbf {\bibinfo {volume} {53}},\ \bibinfo {pages} {182001}
  (\bibinfo {year} {2020})}\BibitemShut {NoStop}%
\bibitem [{\citenamefont {Zhang}\ \emph {et~al.}(2010)\citenamefont {Zhang},
  \citenamefont {Isenhower}, \citenamefont {Gill}, \citenamefont {Walker},\
  and\ \citenamefont {Saffman}}]{zhang2010deterministic}%
  \BibitemOpen
  \bibfield  {author} {\bibinfo {author} {\bibfnamefont {X.-L.}\ \bibnamefont
  {Zhang}}, \bibinfo {author} {\bibfnamefont {L.}~\bibnamefont {Isenhower}},
  \bibinfo {author} {\bibfnamefont {A.}~\bibnamefont {Gill}}, \bibinfo {author}
  {\bibfnamefont {T.-G.}\ \bibnamefont {Walker}},\ and\ \bibinfo {author}
  {\bibfnamefont {M.}~\bibnamefont {Saffman}},\ }\bibfield  {title} {\bibinfo
  {title} {Deterministic entanglement of two neutral atoms via rydberg
  blockade},\ }\href@noop {} {\bibfield  {journal} {\bibinfo  {journal} {Phys.
  Rev. A}\ }\textbf {\bibinfo {volume} {82}},\ \bibinfo {pages} {030306}
  (\bibinfo {year} {2010})}\BibitemShut {NoStop}%
\bibitem [{\citenamefont {Wilk}\ \emph {et~al.}(2010)\citenamefont {Wilk},
  \citenamefont {Ga{\"e}tan}, \citenamefont {Evellin}, \citenamefont {Wolters},
  \citenamefont {Miroshnychenko}, \citenamefont {Grangier},\ and\ \citenamefont
  {Browaeys}}]{wilk2010entanglement}%
  \BibitemOpen
  \bibfield  {author} {\bibinfo {author} {\bibfnamefont {T.}~\bibnamefont
  {Wilk}}, \bibinfo {author} {\bibfnamefont {A.}~\bibnamefont {Ga{\"e}tan}},
  \bibinfo {author} {\bibfnamefont {C.}~\bibnamefont {Evellin}}, \bibinfo
  {author} {\bibfnamefont {J.}~\bibnamefont {Wolters}}, \bibinfo {author}
  {\bibfnamefont {Y.}~\bibnamefont {Miroshnychenko}}, \bibinfo {author}
  {\bibfnamefont {P.}~\bibnamefont {Grangier}},\ and\ \bibinfo {author}
  {\bibfnamefont {A.}~\bibnamefont {Browaeys}},\ }\bibfield  {title} {\bibinfo
  {title} {Entanglement of two individual neutral atoms using rydberg
  blockade},\ }\href@noop {} {\bibfield  {journal} {\bibinfo  {journal} {Phys.
  Rev. Lett.}\ }\textbf {\bibinfo {volume} {104}},\ \bibinfo {pages} {010502}
  (\bibinfo {year} {2010})}\BibitemShut {NoStop}%
\bibitem [{\citenamefont {Levine}\ \emph {et~al.}(2018)\citenamefont {Levine},
  \citenamefont {Keesling}, \citenamefont {Omran}, \citenamefont {Bernien},
  \citenamefont {Schwartz}, \citenamefont {Zibrov}, \citenamefont {Endres},
  \citenamefont {Greiner}, \citenamefont {Vuleti{\'c}},\ and\ \citenamefont
  {Lukin}}]{levine2018high}%
  \BibitemOpen
  \bibfield  {author} {\bibinfo {author} {\bibfnamefont {H.}~\bibnamefont
  {Levine}}, \bibinfo {author} {\bibfnamefont {A.}~\bibnamefont {Keesling}},
  \bibinfo {author} {\bibfnamefont {A.}~\bibnamefont {Omran}}, \bibinfo
  {author} {\bibfnamefont {H.}~\bibnamefont {Bernien}}, \bibinfo {author}
  {\bibfnamefont {S.}~\bibnamefont {Schwartz}}, \bibinfo {author}
  {\bibfnamefont {A.~S.}\ \bibnamefont {Zibrov}}, \bibinfo {author}
  {\bibfnamefont {M.}~\bibnamefont {Endres}}, \bibinfo {author} {\bibfnamefont
  {M.}~\bibnamefont {Greiner}}, \bibinfo {author} {\bibfnamefont
  {V.}~\bibnamefont {Vuleti{\'c}}},\ and\ \bibinfo {author} {\bibfnamefont
  {M.~D.}\ \bibnamefont {Lukin}},\ }\bibfield  {title} {\bibinfo {title}
  {High-fidelity control and entanglement of rydberg-atom qubits},\ }\href@noop
  {} {\bibfield  {journal} {\bibinfo  {journal} {Phys. Rev. Lett.}\ }\textbf
  {\bibinfo {volume} {121}},\ \bibinfo {pages} {123603} (\bibinfo {year}
  {2018})}\BibitemShut {NoStop}%
\bibitem [{\citenamefont {Omran}\ \emph {et~al.}(2019)\citenamefont {Omran},
  \citenamefont {Levine}, \citenamefont {Keesling}, \citenamefont {Semeghini},
  \citenamefont {Wang}, \citenamefont {Ebadi}, \citenamefont {Bernien},
  \citenamefont {Zibrov}, \citenamefont {Pichler}, \citenamefont {Choi} \emph
  {et~al.}}]{omran2019generation}%
  \BibitemOpen
  \bibfield  {author} {\bibinfo {author} {\bibfnamefont {A.}~\bibnamefont
  {Omran}}, \bibinfo {author} {\bibfnamefont {H.}~\bibnamefont {Levine}},
  \bibinfo {author} {\bibfnamefont {A.}~\bibnamefont {Keesling}}, \bibinfo
  {author} {\bibfnamefont {G.}~\bibnamefont {Semeghini}}, \bibinfo {author}
  {\bibfnamefont {T.~T.}\ \bibnamefont {Wang}}, \bibinfo {author}
  {\bibfnamefont {S.}~\bibnamefont {Ebadi}}, \bibinfo {author} {\bibfnamefont
  {H.}~\bibnamefont {Bernien}}, \bibinfo {author} {\bibfnamefont {A.~S.}\
  \bibnamefont {Zibrov}}, \bibinfo {author} {\bibfnamefont {H.}~\bibnamefont
  {Pichler}}, \bibinfo {author} {\bibfnamefont {S.}~\bibnamefont {Choi}}, \emph
  {et~al.},\ }\bibfield  {title} {\bibinfo {title} {Generation and manipulation
  of schr$\backslash$" odinger cat states in rydberg atom arrays},\ }\href@noop
  {} {\bibfield  {journal} {\bibinfo  {journal} {Science}\ }\textbf {\bibinfo
  {volume} {365}},\ \bibinfo {pages} {570} (\bibinfo {year}
  {2019})}\BibitemShut {NoStop}%
\bibitem [{\citenamefont {Graham}\ \emph {et~al.}(2019)\citenamefont {Graham},
  \citenamefont {Kwon}, \citenamefont {Grinkemeyer}, \citenamefont {Marra},
  \citenamefont {Jiang}, \citenamefont {Lichtman}, \citenamefont {Sun},
  \citenamefont {Ebert},\ and\ \citenamefont {Saffman}}]{graham2019rydberg}%
  \BibitemOpen
  \bibfield  {author} {\bibinfo {author} {\bibfnamefont {T.}~\bibnamefont
  {Graham}}, \bibinfo {author} {\bibfnamefont {M.}~\bibnamefont {Kwon}},
  \bibinfo {author} {\bibfnamefont {B.}~\bibnamefont {Grinkemeyer}}, \bibinfo
  {author} {\bibfnamefont {Z.}~\bibnamefont {Marra}}, \bibinfo {author}
  {\bibfnamefont {X.}~\bibnamefont {Jiang}}, \bibinfo {author} {\bibfnamefont
  {M.}~\bibnamefont {Lichtman}}, \bibinfo {author} {\bibfnamefont
  {Y.}~\bibnamefont {Sun}}, \bibinfo {author} {\bibfnamefont {M.}~\bibnamefont
  {Ebert}},\ and\ \bibinfo {author} {\bibfnamefont {M.}~\bibnamefont
  {Saffman}},\ }\bibfield  {title} {\bibinfo {title} {Rydberg mediated
  entanglement in a two-dimensional neutral atom qubit array},\ }\href@noop {}
  {\bibfield  {journal} {\bibinfo  {journal} {arXiv preprint arXiv:1908.06103}\
  } (\bibinfo {year} {2019})}\BibitemShut {NoStop}%
\bibitem [{\citenamefont {Jo}\ \emph {et~al.}(2020)\citenamefont {Jo},
  \citenamefont {Song}, \citenamefont {Kim},\ and\ \citenamefont
  {Ahn}}]{jo2020rydberg}%
  \BibitemOpen
  \bibfield  {author} {\bibinfo {author} {\bibfnamefont {H.}~\bibnamefont
  {Jo}}, \bibinfo {author} {\bibfnamefont {Y.}~\bibnamefont {Song}}, \bibinfo
  {author} {\bibfnamefont {M.}~\bibnamefont {Kim}},\ and\ \bibinfo {author}
  {\bibfnamefont {J.}~\bibnamefont {Ahn}},\ }\bibfield  {title} {\bibinfo
  {title} {Rydberg atom entanglements in the weak coupling regime},\
  }\href@noop {} {\bibfield  {journal} {\bibinfo  {journal} {Phys. Rev. Lett.}\
  }\textbf {\bibinfo {volume} {124}},\ \bibinfo {pages} {033603} (\bibinfo
  {year} {2020})}\BibitemShut {NoStop}%
\bibitem [{\citenamefont {Martin}\ \emph {et~al.}(2021)\citenamefont {Martin},
  \citenamefont {Jau}, \citenamefont {Lee}, \citenamefont {Mitra},
  \citenamefont {Deutsch},\ and\ \citenamefont
  {Biedermann}}]{martin2021molmer}%
  \BibitemOpen
  \bibfield  {author} {\bibinfo {author} {\bibfnamefont {M.~J.}\ \bibnamefont
  {Martin}}, \bibinfo {author} {\bibfnamefont {Y.-Y.}\ \bibnamefont {Jau}},
  \bibinfo {author} {\bibfnamefont {J.}~\bibnamefont {Lee}}, \bibinfo {author}
  {\bibfnamefont {A.}~\bibnamefont {Mitra}}, \bibinfo {author} {\bibfnamefont
  {I.~H.}\ \bibnamefont {Deutsch}},\ and\ \bibinfo {author} {\bibfnamefont
  {G.~W.}\ \bibnamefont {Biedermann}},\ }\bibfield  {title} {\bibinfo {title}
  {A mølmer-sørensen gate with rydberg-dressed atoms},\ }\href@noop {}
  {\bibfield  {journal} {\bibinfo  {journal} {arXiv preprint arXiv:2111.14677}\
  } (\bibinfo {year} {2021})}\BibitemShut {NoStop}%
\bibitem [{\citenamefont {Madjarov}\ \emph {et~al.}(2019)\citenamefont
  {Madjarov}, \citenamefont {Covey}, \citenamefont {Cooper}, \citenamefont
  {Shaw}, \citenamefont {Schkolnik}, \citenamefont {White}, \citenamefont
  {Williams},\ and\ \citenamefont {Endres}}]{madjarov2019strontium}%
  \BibitemOpen
  \bibfield  {author} {\bibinfo {author} {\bibfnamefont {I.}~\bibnamefont
  {Madjarov}}, \bibinfo {author} {\bibfnamefont {J.}~\bibnamefont {Covey}},
  \bibinfo {author} {\bibfnamefont {A.}~\bibnamefont {Cooper}}, \bibinfo
  {author} {\bibfnamefont {A.}~\bibnamefont {Shaw}}, \bibinfo {author}
  {\bibfnamefont {V.}~\bibnamefont {Schkolnik}}, \bibinfo {author}
  {\bibfnamefont {R.}~\bibnamefont {White}}, \bibinfo {author} {\bibfnamefont
  {J.}~\bibnamefont {Williams}},\ and\ \bibinfo {author} {\bibfnamefont
  {M.}~\bibnamefont {Endres}},\ }\bibfield  {title} {\bibinfo {title}
  {Strontium atom arrays: toward rydberg entanglement and optical qubit
  control},\ }\href@noop {} {\bibfield  {journal} {\bibinfo  {journal}
  {Bulletin of the American Physical Society}\ } (\bibinfo {year}
  {2019})}\BibitemShut {NoStop}%
\bibitem [{\citenamefont {Ma}\ \emph {et~al.}(2021)\citenamefont {Ma},
  \citenamefont {Burgers}, \citenamefont {Liu}, \citenamefont {Wilson},
  \citenamefont {Zhang},\ and\ \citenamefont {Thompson}}]{ma2021universal}%
  \BibitemOpen
  \bibfield  {author} {\bibinfo {author} {\bibfnamefont {S.}~\bibnamefont
  {Ma}}, \bibinfo {author} {\bibfnamefont {A.~P.}\ \bibnamefont {Burgers}},
  \bibinfo {author} {\bibfnamefont {G.}~\bibnamefont {Liu}}, \bibinfo {author}
  {\bibfnamefont {J.}~\bibnamefont {Wilson}}, \bibinfo {author} {\bibfnamefont
  {B.}~\bibnamefont {Zhang}},\ and\ \bibinfo {author} {\bibfnamefont {J.~D.}\
  \bibnamefont {Thompson}},\ }\bibfield  {title} {\bibinfo {title} {Universal
  gate operations on nuclear spin qubits in an optical tweezer array of
  ${}^{171}$ yb atoms},\ }\href@noop {} {\bibfield  {journal} {\bibinfo
  {journal} {arXiv preprint arXiv:2112.06799}\ } (\bibinfo {year}
  {2021})}\BibitemShut {NoStop}%
\bibitem [{\citenamefont {Johnson}\ and\ \citenamefont
  {Rolston}(2010)}]{johnson2010interactions}%
  \BibitemOpen
  \bibfield  {author} {\bibinfo {author} {\bibfnamefont {J.-E.}\ \bibnamefont
  {Johnson}}\ and\ \bibinfo {author} {\bibfnamefont {S.-L.}\ \bibnamefont
  {Rolston}},\ }\bibfield  {title} {\bibinfo {title} {Interactions between
  rydberg-dressed atoms},\ }\href@noop {} {\bibfield  {journal} {\bibinfo
  {journal} {Phys. Rev. A}\ }\textbf {\bibinfo {volume} {82}},\ \bibinfo
  {pages} {033412} (\bibinfo {year} {2010})}\BibitemShut {NoStop}%
\bibitem [{\citenamefont {Jau}\ \emph {et~al.}(2016)\citenamefont {Jau},
  \citenamefont {Hankin}, \citenamefont {Keating}, \citenamefont {Deutsch},\
  and\ \citenamefont {Biedermann}}]{jau2016entangling}%
  \BibitemOpen
  \bibfield  {author} {\bibinfo {author} {\bibfnamefont {Y.-Y.}\ \bibnamefont
  {Jau}}, \bibinfo {author} {\bibfnamefont {A.}~\bibnamefont {Hankin}},
  \bibinfo {author} {\bibfnamefont {T.}~\bibnamefont {Keating}}, \bibinfo
  {author} {\bibfnamefont {I.}~\bibnamefont {Deutsch}},\ and\ \bibinfo {author}
  {\bibfnamefont {G.}~\bibnamefont {Biedermann}},\ }\bibfield  {title}
  {\bibinfo {title} {Entangling atomic spins with a rydberg-dressed spin-flip
  blockade},\ }\href@noop {} {\bibfield  {journal} {\bibinfo  {journal} {Nature
  Physics}\ }\textbf {\bibinfo {volume} {12}},\ \bibinfo {pages} {71} (\bibinfo
  {year} {2016})}\BibitemShut {NoStop}%
\bibitem [{\citenamefont {Zhang}\ \emph {et~al.}(2012)\citenamefont {Zhang},
  \citenamefont {Gill}, \citenamefont {Isenhower}, \citenamefont {Walker},\
  and\ \citenamefont {Saffman}}]{zhang2012fidelity}%
  \BibitemOpen
  \bibfield  {author} {\bibinfo {author} {\bibfnamefont {X.}~\bibnamefont
  {Zhang}}, \bibinfo {author} {\bibfnamefont {A.}~\bibnamefont {Gill}},
  \bibinfo {author} {\bibfnamefont {L.}~\bibnamefont {Isenhower}}, \bibinfo
  {author} {\bibfnamefont {T.}~\bibnamefont {Walker}},\ and\ \bibinfo {author}
  {\bibfnamefont {M.}~\bibnamefont {Saffman}},\ }\bibfield  {title} {\bibinfo
  {title} {Fidelity of a rydberg-blockade quantum gate from simulated quantum
  process tomography},\ }\href@noop {} {\bibfield  {journal} {\bibinfo
  {journal} {Phys. Rev. A}\ }\textbf {\bibinfo {volume} {85}},\ \bibinfo
  {pages} {042310} (\bibinfo {year} {2012})}\BibitemShut {NoStop}%
\bibitem [{\citenamefont {de~L{\'e}s{\'e}leuc}\ \emph
  {et~al.}(2018)\citenamefont {de~L{\'e}s{\'e}leuc}, \citenamefont {Barredo},
  \citenamefont {Lienhard}, \citenamefont {Browaeys},\ and\ \citenamefont
  {Lahaye}}]{de2018analysis}%
  \BibitemOpen
  \bibfield  {author} {\bibinfo {author} {\bibfnamefont {S.}~\bibnamefont
  {de~L{\'e}s{\'e}leuc}}, \bibinfo {author} {\bibfnamefont {D.}~\bibnamefont
  {Barredo}}, \bibinfo {author} {\bibfnamefont {V.}~\bibnamefont {Lienhard}},
  \bibinfo {author} {\bibfnamefont {A.}~\bibnamefont {Browaeys}},\ and\
  \bibinfo {author} {\bibfnamefont {T.}~\bibnamefont {Lahaye}},\ }\bibfield
  {title} {\bibinfo {title} {Analysis of imperfections in the coherent optical
  excitation of single atoms to rydberg states},\ }\href@noop {} {\bibfield
  {journal} {\bibinfo  {journal} {Phys. Rev. A}\ }\textbf {\bibinfo {volume}
  {97}},\ \bibinfo {pages} {053803} (\bibinfo {year} {2018})}\BibitemShut
  {NoStop}%
\bibitem [{\citenamefont {Keating}\ \emph {et~al.}(2013)\citenamefont
  {Keating}, \citenamefont {Goyal}, \citenamefont {Jau}, \citenamefont
  {Biedermann}, \citenamefont {Landahl},\ and\ \citenamefont
  {Deutsch}}]{keating2013adiabatic}%
  \BibitemOpen
  \bibfield  {author} {\bibinfo {author} {\bibfnamefont {T.}~\bibnamefont
  {Keating}}, \bibinfo {author} {\bibfnamefont {K.}~\bibnamefont {Goyal}},
  \bibinfo {author} {\bibfnamefont {Y.-Y.}\ \bibnamefont {Jau}}, \bibinfo
  {author} {\bibfnamefont {G.~W.}\ \bibnamefont {Biedermann}}, \bibinfo
  {author} {\bibfnamefont {A.~J.}\ \bibnamefont {Landahl}},\ and\ \bibinfo
  {author} {\bibfnamefont {I.~H.}\ \bibnamefont {Deutsch}},\ }\bibfield
  {title} {\bibinfo {title} {Adiabatic quantum computation with rydberg-dressed
  atoms},\ }\href@noop {} {\bibfield  {journal} {\bibinfo  {journal} {Phys.
  Rev. A}\ }\textbf {\bibinfo {volume} {87}},\ \bibinfo {pages} {052314}
  (\bibinfo {year} {2013})}\BibitemShut {NoStop}%
\bibitem [{\citenamefont {Wesenberg}\ \emph {et~al.}(2007)\citenamefont
  {Wesenberg}, \citenamefont {M{\o}lmer}, \citenamefont {Rippe},\ and\
  \citenamefont {Kr{\"o}ll}}]{wesenberg2007scalable}%
  \BibitemOpen
  \bibfield  {author} {\bibinfo {author} {\bibfnamefont {J.~H.}\ \bibnamefont
  {Wesenberg}}, \bibinfo {author} {\bibfnamefont {K.}~\bibnamefont
  {M{\o}lmer}}, \bibinfo {author} {\bibfnamefont {L.}~\bibnamefont {Rippe}},\
  and\ \bibinfo {author} {\bibfnamefont {S.}~\bibnamefont {Kr{\"o}ll}},\
  }\bibfield  {title} {\bibinfo {title} {Scalable designs for quantum computing
  with rare-earth-ion-doped crystals},\ }\href@noop {} {\bibfield  {journal}
  {\bibinfo  {journal} {Phys. Rev. A}\ }\textbf {\bibinfo {volume} {75}},\
  \bibinfo {pages} {012304} (\bibinfo {year} {2007})}\BibitemShut {NoStop}%
\bibitem [{\citenamefont {Sa{\ss}mannshausen}\ and\ \citenamefont
  {Deiglmayr}(2016)}]{sassmannshausen2016observation}%
  \BibitemOpen
  \bibfield  {author} {\bibinfo {author} {\bibfnamefont {H.}~\bibnamefont
  {Sa{\ss}mannshausen}}\ and\ \bibinfo {author} {\bibfnamefont
  {J.}~\bibnamefont {Deiglmayr}},\ }\bibfield  {title} {\bibinfo {title}
  {Observation of rydberg-atom macrodimers: Micrometer-sized diatomic
  molecules},\ }\href@noop {} {\bibfield  {journal} {\bibinfo  {journal} {Phys.
  Rev. Lett.}\ }\textbf {\bibinfo {volume} {117}},\ \bibinfo {pages} {083401}
  (\bibinfo {year} {2016})}\BibitemShut {NoStop}%
\bibitem [{\citenamefont {Hollerith}\ \emph {et~al.}(2021)\citenamefont
  {Hollerith}, \citenamefont {Srakaew}, \citenamefont {Wei}, \citenamefont
  {Rubio-Abadal}, \citenamefont {Adler}, \citenamefont {Weckesser},
  \citenamefont {Kruckenhauser}, \citenamefont {Walther}, \citenamefont {van
  Bijnen}, \citenamefont {Rui} \emph {et~al.}}]{hollerith2021realizing}%
  \BibitemOpen
  \bibfield  {author} {\bibinfo {author} {\bibfnamefont {S.}~\bibnamefont
  {Hollerith}}, \bibinfo {author} {\bibfnamefont {K.}~\bibnamefont {Srakaew}},
  \bibinfo {author} {\bibfnamefont {D.}~\bibnamefont {Wei}}, \bibinfo {author}
  {\bibfnamefont {A.}~\bibnamefont {Rubio-Abadal}}, \bibinfo {author}
  {\bibfnamefont {D.}~\bibnamefont {Adler}}, \bibinfo {author} {\bibfnamefont
  {P.}~\bibnamefont {Weckesser}}, \bibinfo {author} {\bibfnamefont
  {A.}~\bibnamefont {Kruckenhauser}}, \bibinfo {author} {\bibfnamefont
  {V.}~\bibnamefont {Walther}}, \bibinfo {author} {\bibfnamefont
  {R.}~\bibnamefont {van Bijnen}}, \bibinfo {author} {\bibfnamefont
  {J.}~\bibnamefont {Rui}}, \emph {et~al.},\ }\bibfield  {title} {\bibinfo
  {title} {Realizing distance-selective interactions in a rydberg-dressed atom
  array},\ }\href@noop {} {\bibfield  {journal} {\bibinfo  {journal} {arXiv
  preprint arXiv:2110.10125}\ } (\bibinfo {year} {2021})}\BibitemShut {NoStop}%
\bibitem [{\citenamefont {Lee}\ \emph {et~al.}(2017)\citenamefont {Lee},
  \citenamefont {Martin}, \citenamefont {Jau}, \citenamefont {Keating},
  \citenamefont {Deutsch},\ and\ \citenamefont
  {Biedermann}}]{lee2017demonstration}%
  \BibitemOpen
  \bibfield  {author} {\bibinfo {author} {\bibfnamefont {J.}~\bibnamefont
  {Lee}}, \bibinfo {author} {\bibfnamefont {M.~J.}\ \bibnamefont {Martin}},
  \bibinfo {author} {\bibfnamefont {Y.-Y.}\ \bibnamefont {Jau}}, \bibinfo
  {author} {\bibfnamefont {T.}~\bibnamefont {Keating}}, \bibinfo {author}
  {\bibfnamefont {I.~H.}\ \bibnamefont {Deutsch}},\ and\ \bibinfo {author}
  {\bibfnamefont {G.~W.}\ \bibnamefont {Biedermann}},\ }\bibfield  {title}
  {\bibinfo {title} {Demonstration of the jaynes-cummings ladder with
  rydberg-dressed atoms},\ }\href@noop {} {\bibfield  {journal} {\bibinfo
  {journal} {Phys. Rev. A}\ }\textbf {\bibinfo {volume} {95}},\ \bibinfo
  {pages} {041801(R)} (\bibinfo {year} {2017})}\BibitemShut {NoStop}%
\bibitem [{\citenamefont {Nielsen}\ \emph {et~al.}(2003)\citenamefont
  {Nielsen}, \citenamefont {Dawson}, \citenamefont {Dodd}, \citenamefont
  {Gilchrist}, \citenamefont {Mortimer}, \citenamefont {Osborne}, \citenamefont
  {Bremner}, \citenamefont {Harrow},\ and\ \citenamefont
  {Hines}}]{nielsen2003quantum}%
  \BibitemOpen
  \bibfield  {author} {\bibinfo {author} {\bibfnamefont {M.~A.}\ \bibnamefont
  {Nielsen}}, \bibinfo {author} {\bibfnamefont {C.~M.}\ \bibnamefont {Dawson}},
  \bibinfo {author} {\bibfnamefont {J.~L.}\ \bibnamefont {Dodd}}, \bibinfo
  {author} {\bibfnamefont {A.}~\bibnamefont {Gilchrist}}, \bibinfo {author}
  {\bibfnamefont {D.}~\bibnamefont {Mortimer}}, \bibinfo {author}
  {\bibfnamefont {T.~J.}\ \bibnamefont {Osborne}}, \bibinfo {author}
  {\bibfnamefont {M.~J.}\ \bibnamefont {Bremner}}, \bibinfo {author}
  {\bibfnamefont {A.~W.}\ \bibnamefont {Harrow}},\ and\ \bibinfo {author}
  {\bibfnamefont {A.}~\bibnamefont {Hines}},\ }\bibfield  {title} {\bibinfo
  {title} {Quantum dynamics as a physical resource},\ }\href@noop {} {\bibfield
   {journal} {\bibinfo  {journal} {Phys. Rev. A}\ }\textbf {\bibinfo {volume}
  {67}},\ \bibinfo {pages} {052301} (\bibinfo {year} {2003})}\BibitemShut
  {NoStop}%
\bibitem [{\citenamefont {Zhang}\ \emph {et~al.}(2003)\citenamefont {Zhang},
  \citenamefont {Vala}, \citenamefont {Sastry},\ and\ \citenamefont
  {Whaley}}]{zhang2003geometric}%
  \BibitemOpen
  \bibfield  {author} {\bibinfo {author} {\bibfnamefont {J.}~\bibnamefont
  {Zhang}}, \bibinfo {author} {\bibfnamefont {J.}~\bibnamefont {Vala}},
  \bibinfo {author} {\bibfnamefont {S.}~\bibnamefont {Sastry}},\ and\ \bibinfo
  {author} {\bibfnamefont {K.~B.}\ \bibnamefont {Whaley}},\ }\bibfield  {title}
  {\bibinfo {title} {Geometric theory of nonlocal two-qubit operations},\
  }\href@noop {} {\bibfield  {journal} {\bibinfo  {journal} {Phys. Rev. A}\
  }\textbf {\bibinfo {volume} {67}},\ \bibinfo {pages} {042313} (\bibinfo
  {year} {2003})}\BibitemShut {NoStop}%
\end{thebibliography}%

\end{document}